\pdfoutput=1
\documentclass[11pt,a4paper,UKenglish]{article}

\usepackage{jheppub}

\usepackage{atlaspackage}

\usepackage{atlasphysics}

\graphicspath{{./}{./}}

\hypersetup{pdftitle={ATLAS draft},pdfauthor={The ATLAS Collaboration}}

\usepackage[preprint]{atlascover}

\AtlasTitle{Search for low-scale gravity signatures in multi-jet final states
with the ATLAS detector at $\sqrt{s} = 8\TeV$}
\AtlasVersion{2.4}

\AtlasRefCode{EXOT-2014-02} 

\AtlasNote{ATLAS-COM-PHYS-2014-674}

\AtlasJournal{Journal of High Energy Physics}

\AtlasAbstract{%
A search for evidence of physics beyond the Standard Model in final states
with multiple high-transverse-momentum jets is performed using \lumi{} of
proton--proton collision data at $\sqrt{s} = 8$~TeV recorded by the ATLAS
detector at the LHC. 
No significant excess of events beyond Standard Model expectations is
observed, and upper limits on the visible cross sections for non-Standard
Model production of multi-jet final states are set.
A wide variety of models for black hole and string ball production
and decay are considered, and the upper limit on the cross section times
acceptance is as low as 0.16~fb at the 95\% confidence level.  
For these models, excluded regions are also given as function of the
main model parameters.  

}

\AtlasCoverSupportingNote{ATL-COM-PHYS-2014-674}{https://cds.cern.ch/record/1710778}
\AtlasCoverCommentsDeadline{12 March 2015}

\AtlasCoverAnalysisTeam{Marco~Cardaci, Patrick~Czodrowski~(*), Douglas~M.~Gingrich~(*),
Daniel~Turgeman, Kuhan~Wang~(*)}

\AtlasCoverEdBoardMember{Ian~Hinchliffe~(*), Mario~Martinez, Josh~McFayden,
Domizia~Orestano}

\AtlasCoverEgroupEditors{atlas-exot-2014-02-editors@cern.ch}

\AtlasCoverEgroupEdBoard{atlas-exot-2014-02-editorial-board@cern.ch}

\PreprintIdNumber{CERN-PH-EP-2015-066}

\usepackage{multijet-defs}

\DeclareGraphicsExtensions{.pdf}

\begin{document}

\title{\AtlasTitleText}

\author{The ATLAS Collaboration}

\begin{abstract}
\AtlasAbstractText
\end{abstract}

\maketitle

\section{Introduction}
\label{sec:intro}

Most models of low-scale gravity allow the production of non-perturbative
gravitational states, such as micro black holes and string balls
(highly excited string states) at Large Hadron Collider (LHC) collision
energies~\cite{ArkaniHamed:1998rs,Antoniadis:1998ig,Randall:1999ee,Randall:1999vf}.
This is due to the fundamental gravitational scale being comparable to
the electroweak scale ($M_\mathrm{EW}$) in these gravity models.
If black holes or string balls are produced at the LHC with masses much
higher than this fundamental gravitational scale, they behave as
classical thermal states and decay to a relatively large number of
high-transverse-momentum (high-\pT) particles.
One of the predictions of these models is the expectation that particles are
emitted from black holes primarily according to the number of Standard
Model (SM) degrees of freedom (number of charge, spin, flavour, and
colour states).

To identify high-\pT, high-multiplicity final states resulting from
high-mass objects, a suitable variable is the scalar sum of the \pT{} of
the jets in the event, \HT{}.
A low-\HT{} control region is defined where the background is expected
to dominate over any possible new physics signal.
A fit-based technique is used to extrapolate from the control region to a
high-\HT{} signal region to estimate the amount of SM background.

This paper is organised as follows.
The phenomenology of low-scale gravity relevant to the search is briefly
described in section~\ref{sec:theory}.
In section~\ref{sec:atlas}, the main components of the ATLAS detector are
summarised.
The Monte Carlo (MC) simulated samples used for the analysis are presented in
section~\ref{sec:mc}.
In section~\ref{sec:data}, the trigger and event selection are
described.
The characterisation of the data and the method used in the search are
given in section~\ref{sec:method}.
Section~\ref{sec:sys} describes the systematic uncertainties, and the
resulting limits are given in section~\ref{sec:result}.
Finally, conclusions are stated in section~\ref{sec:conclusion}.

\section{Theoretical background and previous results}
\label{sec:theory}

Understanding quantum gravity is one of the main challenges of modern
physics.
The hierarchy problem (the relative weakness of gravity compared to the
electroweak interaction) may be key to that understanding.
Two main paradigms for models involving extra dimensions have
been formulated: the Arkani-Hamed, Dimopoulos, Dvali (ADD)
proposal~\cite{ArkaniHamed:1998rs,Antoniadis:1998ig} involving large
extra dimensions; and a five-dimensional model with a single highly warped
anti-de Sitter space~\cite{Randall:1999ee,Randall:1999vf}.
These models have our (3+1)-dimensional world residing on a
brane, which is embedded in a (4+$n$)-dimensional bulk with $n$
extra dimensions.
The effective strength of the gravitational interaction inside the brane
is weakened by the large volume of the extra dimensions or red-shifted
by the warp factor along the extra dimension.
This weakening of the gravitational strength results in a diminished
effective Planck scale $\MD$ in the (4+$n$)-dimensional world, relative
to the familiar Planck scale $M_\mathrm{Pl}$.
In the ADD model, there are a number $n > 1$
additional flat extra dimensions, and $\MD$ is determined by the
volume and shape of the extra dimensions.

If $\MD \sim M_\mathrm{EW}$, several low-scale gravitational signatures
may be probed in collider physics experiments.
Some of the most interesting are the possible existence of
non-perturbative gravitational states such as black
holes~\cite{ArkaniHamed:1998rs,Antoniadis:1998ig,Randall:1999ee,Randall:1999vf},
string balls~\cite{Dimopoulos:2001qe}  (in the context of
weakly coupling string theory), and higher-dimensional branes.

Within the context of the ADD model, experimental lower limits on the
value of $\MD$~\cite{Gingrich:2012vs} were obtained from experiments at
LEP and the Tevatron~\cite{Aaltonen:2008hh,Abazov:2008kp}, as well as at
ATLAS~\cite{Aad:2015zva} and CMS~\cite{Chatrchyan:2012me}, by searching
for the production of the heavy Kaluza--Klein gravitons associated with
the extra dimensions.
The most stringent limits come from the LHC
analyses~\cite{Chatrchyan:2012me,Aad:2015zva} that search for
non-interacting gravitons recoiling against a single jet,
and range from $\MD > 3.1\,\TeV$, for $n = 6$, to $\MD > 5.2\,\TeV$, for
$n = 2$.
Several searches for black holes and string balls are also performed by
ATLAS~\cite{Aad:2011bw,Aad:2012ic,Aad:2013lna,Aad:2014gka} and
CMS~\cite{Khachatryan:2010wx,Chatrchyan:2012taa,Chatrchyan:2013xva}.

In proton--proton collisions with centre-of-mass energy $\sqrt{s}$,
classical black holes form when the impact parameter between two colliding
partons, with centre-of-mass energy $\sqrt{\hat{s}}$, is less than twice
the gravitational radius $r_\mathrm{g}$ of a black hole of mass equal to
$\sqrt{\hat{s}}$~\cite{Dimopoulos:2001hw,Giddings:2001bu}.
Black holes are assumed to be produced over a continuous range of masses
above a certain threshold $\Mth \gtrsim \MD$ up to $\sqrt{s}$.
Semi-classical approximations used in the modelling are valid for
masses only well above $\MD$, motivating the use of a minimal
threshold \Mth{} to remove contributions where the modelling is not
reliable.

Most low-scale gravity models assume classical general relativity to
predict the production cross section for black holes ($\sigma \sim \pi
r_\mathrm{g}^2$) and string balls, and use semi-classical Hawking
evaporation (a completely thermal process due to quantum effects) to
describe their decay~\cite{Hawking75}.
The decay process is described by black-body radiation at the Hawking
temperature (Hagedorn temperature for string balls) with the
expectation that the radiated particle species are produced according to
the number of SM degrees of freedom and are not affected by the
strengths of the SM forces.
The emissions are modified by spin-dependent quantum statistics given by
the Fermi--Dirac or Bose--Einstein distributions.
In addition, the emissions are modified by gravitational transmission
factors~\cite{Hawking75} (gray-body factors), which depend on the spin of
the emitted particle, as well as the angular momentum of the black hole,
and can be sizeable for vector particle emission from rotating black holes.
Once black holes are produced, they evaporate causing their mass to be
reduced with each emitted particle.
In the context of weakly coupled string theory, black holes transition
to string balls at a minimum black hole mass $M_\mathrm{min} \sim
M_\mathrm{s}/g_\mathrm{s}^2$, where $M_\mathrm{s}$ is the string scale
and $g_\mathrm{s}$ is the string coupling
constant~\cite{Dimopoulos:2001qe,Gingrich:2008di}.
When the black hole mass is reduced to approximately \MD{} (or
$M_\mathrm{s}$ for string balls), the black hole is said to be in a
remnant state, which is expected to only be describable by a theory of
quantum gravity.
This study only considers unstable black hole remnants, and black holes and string balls that are short lived.

The production and decay of black holes and string balls lead to final
states distinguished by a high multiplicity of high-\pT{} particles,
consisting mostly of jets arising from quark and gluon emission.
Since black hole decay is considered to be a stochastic process, a
different number of particles, and thus jets, can be emitted from black
holes with identical kinematics.

\section{ATLAS detector}
\label{sec:atlas}

The ATLAS experiment~\cite{Aad:2008zzm} is a multi-purpose particle
physics detector with a forward-backward symmetric cylindrical
geometry and nearly $4\pi$ coverage in solid angle.\footnote{The ATLAS detector uses
a right-handed coordinate system with its origin at the nominal
interaction point in the centre of the detector and the $z$-axis along
the beam direction.
The $x$-axis points toward the centre of the LHC ring, and the $y$-axis
points upward.
Cylindrical coordinates $(r,\phi)$ are used in the transverse
plane, $\phi$ being the azimuthal angle around the beam pipe.
The pseudorapidity $\eta$ is defined in terms of the polar angle
$\theta$ by $\eta \equiv -\ln[\tan(\theta/2)]$.}
The layout of the detector is dominated by four superconducting
magnet systems, which comprise a thin solenoid surrounding inner
tracking detectors and three large toroids, each consisting of eight
coils.
The inner detector consists of a silicon pixel detector, a silicon
microstrip detector, and a transition radiation tracker, with a combined
coverage up to $|\eta| = 2.5$.
In the pseudorapidity region $|\eta| < 3.2$, liquid-argon (LAr)
electromagnetic (EM) sampling calorimeters are used.
An iron/scintillator tile calorimeter provides hadronic coverage
over $|\eta| < 1.7$.
The end-cap and forward regions, spanning $\HTmin < |\eta| < 4.9$, are
instrumented with LAr calorimetry for EM and hadronic measurements.
The muon spectrometer surrounds these, and comprises a system of
precision tracking and trigger chambers.
A three-level trigger system is used to select interesting
events~\cite{Aad:2012xs}.
The Level-1 trigger is implemented in hardware and uses a subset of
detector information to reduce the event rate to at most 75~kHz.
This is followed by two software-based trigger levels which together
reduce the event rate to about 300~Hz.

\section{Monte Carlo simulation}
\label{sec:mc}

All background estimates in this analysis are derived from data.
However, SM MC simulated events are used to estimate the
relative
background contributions from different processes expected in the data sample,
and to develop and validate the analysis methods.

The dominant background in the search region consists of QCD multi-jet events,
 with small contributions from top quark pair production
($\ttbar$), $\gamma$+jets, $W$+jets, and $Z$+jets.
Single-top-quark and diboson processes contribute negligibly to the
selected samples.
The baseline samples of inclusive jets are generated using
\pythia~8.160~\cite{Pythia:2007} implementing LO perturbative QCD matrix
elements for $2\rightarrow 2$ processes and \pT-ordered parton showers
calculated in a leading-logarithmic approximation.
The ATLAS AU2 set of MC parameters (tune)~\cite{atlas:pythia8tunes} and
the CT10~\cite{Lai:2010vv} PDFs are used with these samples.
\herwigpp~2.6.3~\cite{Bahr:2008pv} dijet samples with the ATLAS EE3 tune
and CTEQ6L1~\cite{Pumplin:2002vw} PDFs, and
\alpgen~2.14~\cite{Mangano:2002ea} multi-jet samples hadronised with
\pythia~6.427 with the ATLAS Perugia 2001C tune and the CTEQ6L1 PDFs are used
for comparisons.
The \ttbar{}, $\gamma$+jets, $W$+jets, and $Z$+jets samples are
generated using \sherpa~1.4.0~\cite{Gleisberg:2008ta} with CT10 PDFs.
All MC simulated background samples (except \alpgen) are using the full
GEANT4~\cite{Agostinelli2003250} simulation.

Signal acceptances are determined using MC simulated events.
Signal samples are generated using the MC event generators
\charybdis~1.0.2~\cite{Frost:2009cf} and
\blackmax~2.02.0~\cite{Dai:2007ki}.
Two generators are used since they model the remnant decay slightly
differently and neither implements all the models considered in this
analysis.
\charybdis is used to produce samples for non-rotating, rotating, and
low-multiplicity remnant black holes, and for an initial-state graviton
radiation model.
\blackmax is used to produce samples for non-rotating and rotating
black holes, and for final-state graviton emission and initial-state
photon radiation models.
The initial-state radiation is modelled to occur after, rather than
before, black hole formation.
In addition, \charybdis is used to produce non-rotating and rotating
string ball samples.
Both generators use a leading-order parton distribution function (PDF)
MSTW2008~\cite{Martin:2009iq}, the ATLAS AU2 tune, and the \pythia~8.165
generator for fragmentation.
The most important parameters that have significant effects on
black hole production are \Mth, $\MD$ ($M_\mathrm{s}$ for string balls),
and $n$.
Signal samples are produced for many values of these parameters.
The MC simulated signal samples are passed through a fast simulation of
the ATLAS detector~\cite{AFII}.
The fast simulation uses a parameterised response of the calorimeters,
and GEANT4 for the other parts of the ATLAS detector.
The difference in signal yield with respect to a full GEANT4 simulation
of the ATLAS detector~\cite{ATLASSimInfra} is negligible.

Additional proton--proton collisions are modelled by overlaying minimum
bias events on the simulated signal and background events according to
the luminosity profile of the recorded data.
The MC simulated events are reconstructed and analysed with
the same procedures as used on data.

\section{Trigger and data selection}
\label{sec:data}

The data used in this analysis were recorded in 2012, with the LHC
operating at a centre-of-mass energy of $\sqrt{s} = 8\TeV$.
All detector elements are required to be fully operational, and a total
integrated luminosity of \lumi{} is used in this analysis with a
luminosity uncertainty of 2.8\%.
It is derived following the same methodology as that detailed in
ref.~\cite{Aad:2013ucp}.

The events used in this search are selected using a high-\HT{} trigger,
which requires at least one jet of hadrons with $\pT > 170\GeV$ and a
high scalar sum of transverse momentum of all the jets in the event.
The trigger is fully efficient if the event has $\HT > 1.2\TeV$, as
required in this analysis.

Events are required to have a primary vertex with at least two
associated tracks with \pT{} above $400\MeV$.
The primary vertex assigned to the hard scattering collision is the one
with the highest $\sum_\mathrm{track} \pT^2$, where the scalar sum of
track $\pT^2$ is taken over all tracks associated with that vertex.

Since black holes and string balls are expected to decay predominantly
to quarks and gluons, the search is simplified by considering only jets.
The analysis uses jets of hadrons, as well as misidentified jets from
photons, electrons, and $\tau$ leptons.
The incorrect calibration of photons, electrons, and $\tau$ leptons
using the hadronic energy calibration leads to small energy shifts for
these particles, but since a particle of this type is expected to occur
in less than 0.6\% (as determined from simulation studies) of the events
in the data sample, they do not contribute significantly to the
resolution of global quantities.

The anti-$k_t$ algorithm~\cite{Cacciari:2008gp} is used for jet finding, with
a radius parameter $R = 0.4$.
The inputs to the jet reconstruction are three-dimensional
topo-clusters~\cite{Lampl}.
This method first clusters together topologically connected calorimeter
cells and then classifies these clusters as either electromagnetic or
hadronic.
The classification uses a local cluster weighting calibration scheme
based on cell-energy density and longitudinal depth within the
calorimeter.
Based on this classification, energy corrections described in
ref.~\cite{Aad:2014bia} are applied.
Furthermore, jets are corrected for pile-up.
The jets are required to have $\pT > 50\GeV$ and $|\eta| < 2.8$ in this
analysis.

\section{Background estimation method}
\label{sec:method}

\begin{figure}[htb]
\begin{center}
\includegraphics[width=0.40\textwidth]{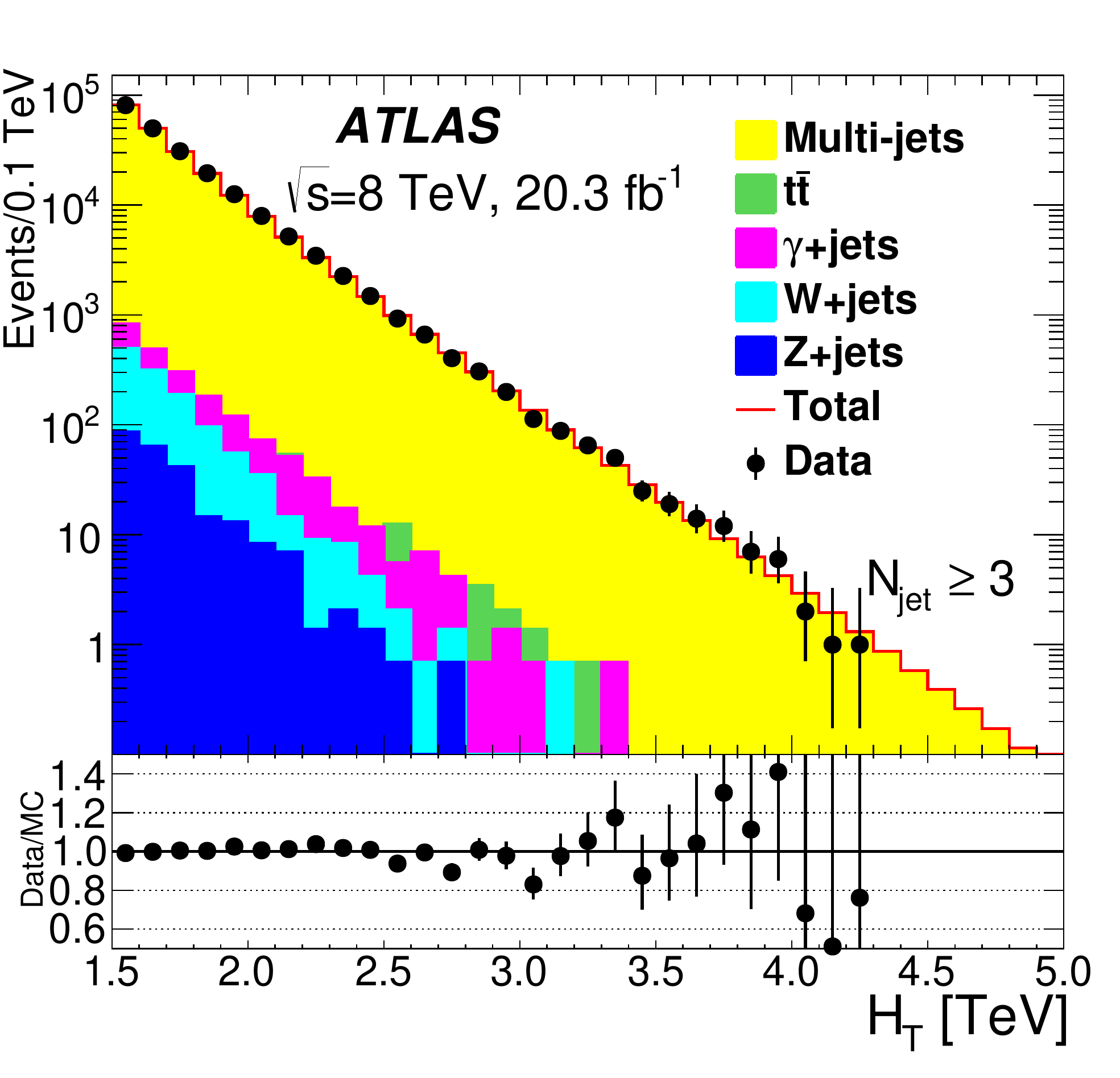}
\includegraphics[width=0.40\textwidth]{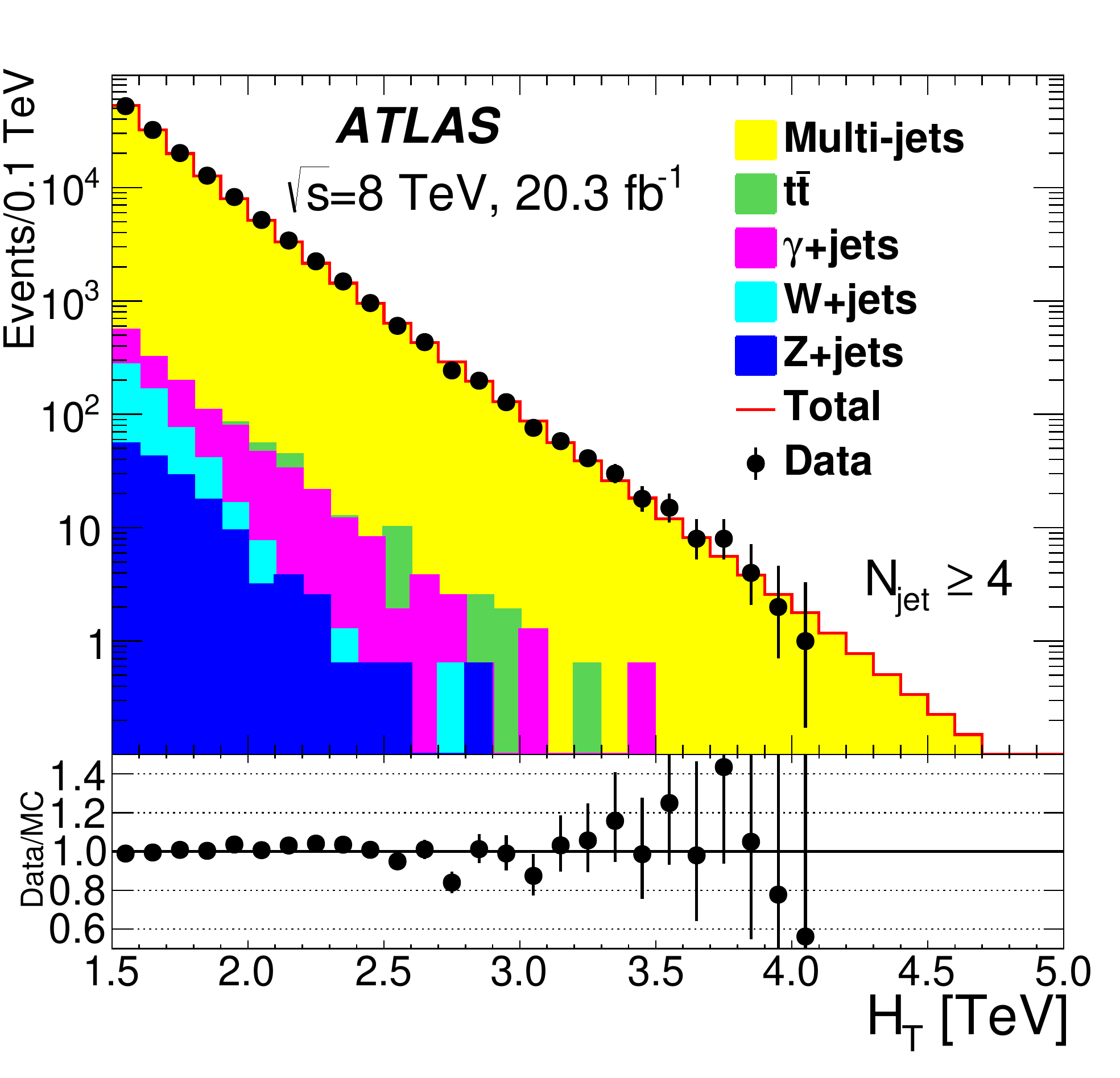}
\includegraphics[width=0.40\textwidth]{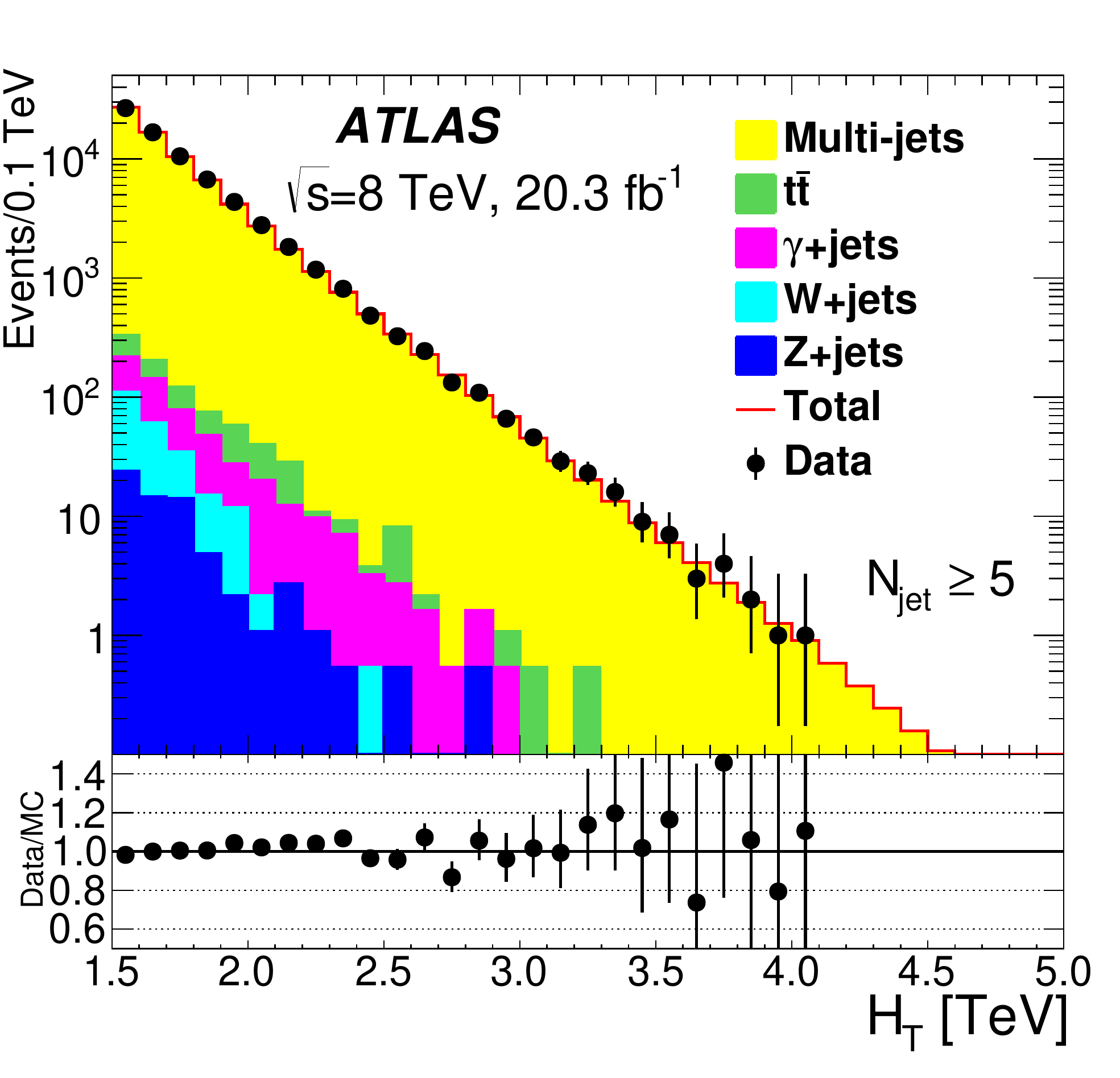}
\includegraphics[width=0.40\textwidth]{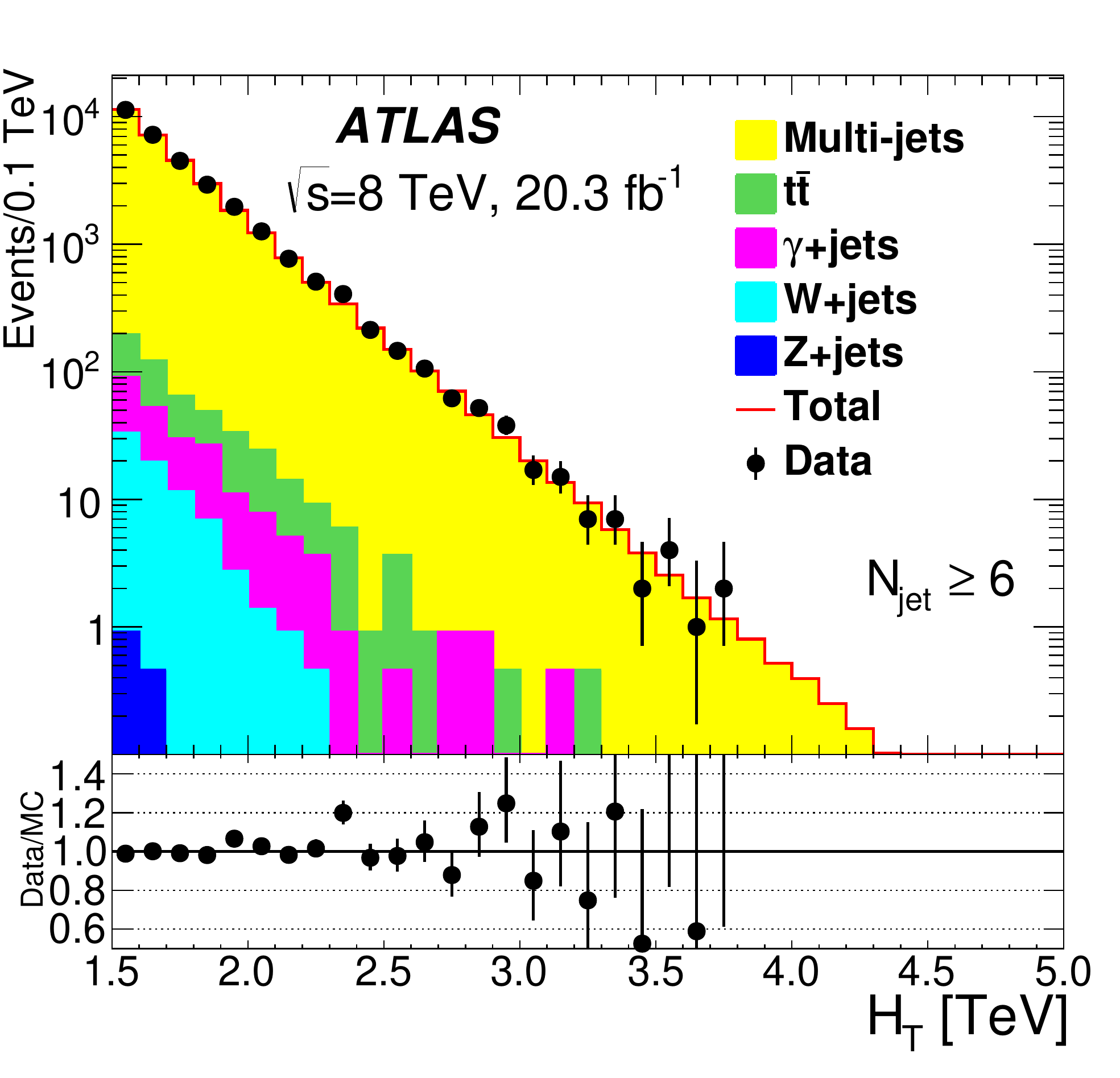}
\includegraphics[width=0.40\textwidth]{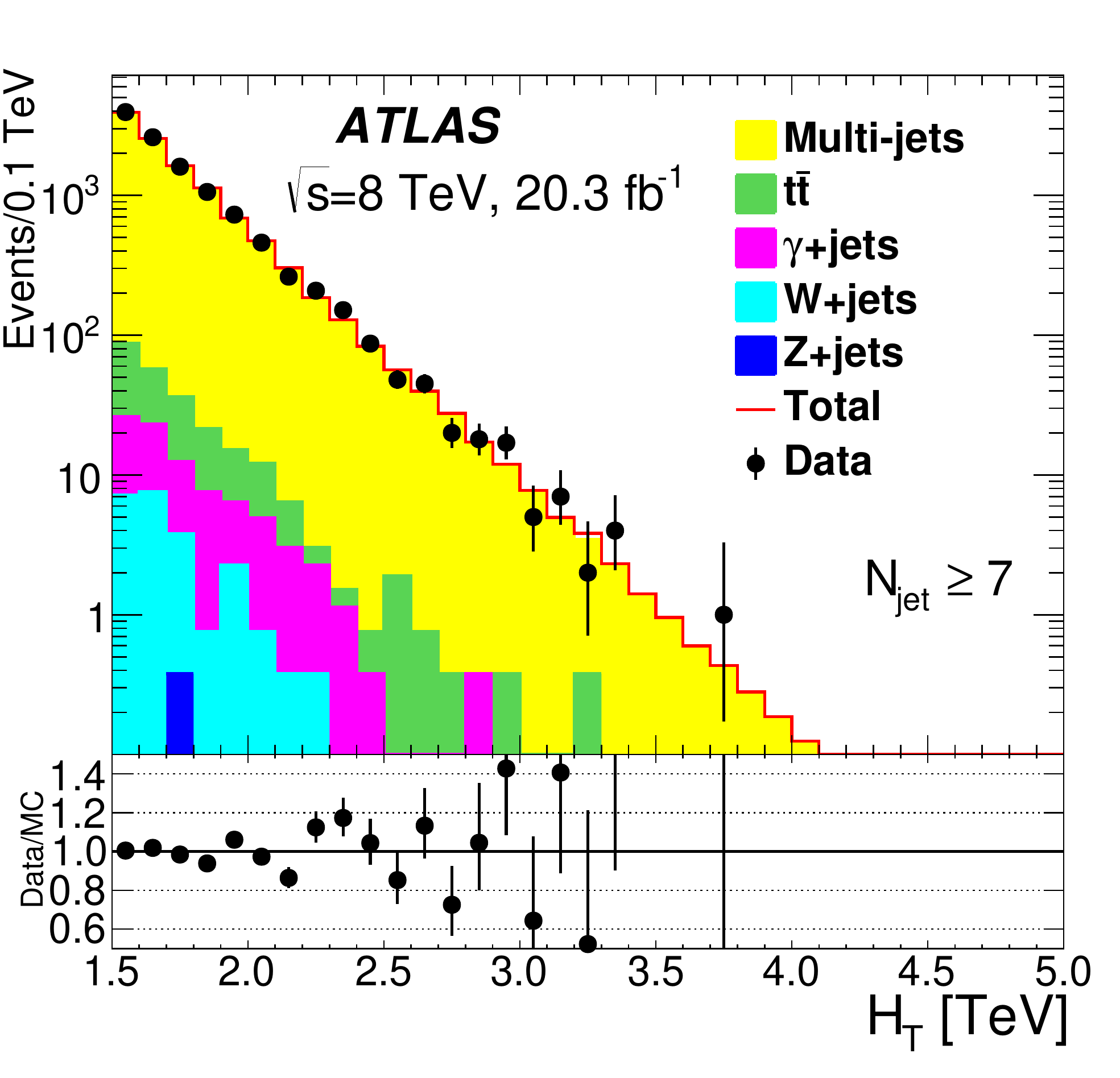}
\includegraphics[width=0.40\textwidth]{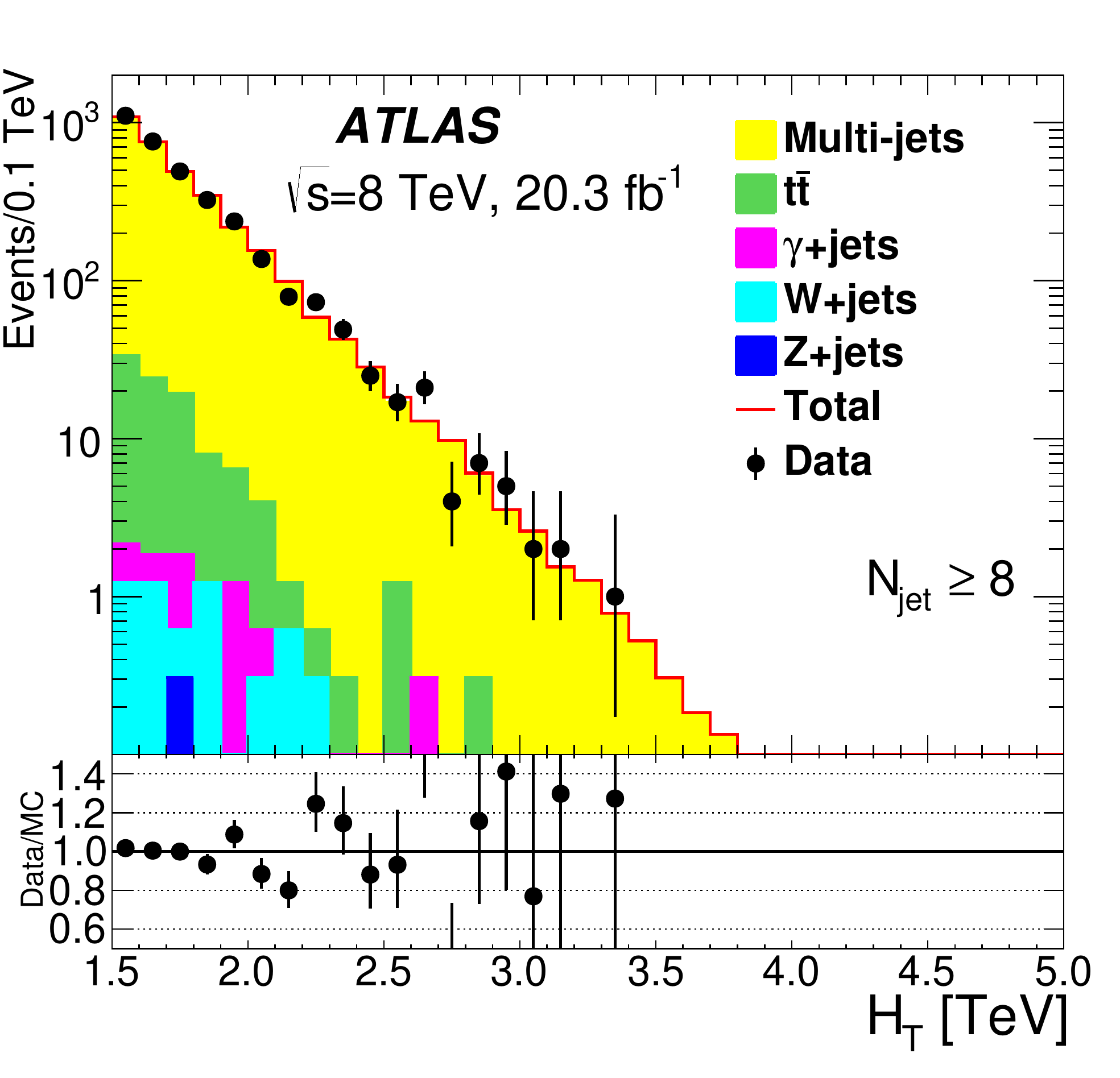}
\caption{Distributions of the scalar sum of the \pT{} of all jets in the
event, \HT, for different inclusive jet multiplicities \NJ{} for \lumi{}
of collision data and MC simulations of SM processes.
The uncertainties on the data and ratio points are due to the statistical
uncertainty of the data only.
The SM contributions to the background are normalised relative to their
nominal cross sections and then the total background is normalised to
the number of data events in the region $\HTmin < \HT < \HTmax\TeV$ for each
inclusive jet multiplicity.}
\label{fig1}
\end{center}
\end{figure}

Events are selected if they pass the high-\HT{} trigger and have $\HT >
1.5\TeV$.
The discriminating variable chosen for this analysis is \HT{}.
Figure~\ref{fig1} shows the \HT{} distributions for different inclusive
jet multiplicities.
Data as well as MC simulations of the most significant SM contributions to the
\HT{} distributions are shown.
The different MC contributions are first weighted according to their
cross sections, and then the total SM contribution is normalised to the
number of data events in the region $\HTmin < \HT < \HTmax\TeV$ for each
inclusive jet multiplicity.
As can be seen in figure~\ref{fig1}, the expected SM background is
dominated by QCD jet production.
The rest of the background processes contribute less than 3\% to the total
background, and therefore the other contributions are neglected in what
follows.
Figure~\ref{fig1} also shows good agreement between SM expectations and
data, which is quantified in section~\ref{sec:result}.

Each event is characterised by the number of jets \NJ{} and by the value
of \HT.
The $(\NJ, \HT)$-variable space is divided into two exclusive \HT{}
regions for the search, which are each further divided by inclusive jet
multiplicity \NJ.
The two exclusive \HT{} regions are defined as a control region ($\HTmin
< \HT < \HTmax\TeV$) and as a signal region ($\HT > 3.0\TeV$), for each
inclusive jet multiplicity.
The control region is utilised to fit the \HT{} distribution, as no
resonances or threshold enhancements above SM processes were observed in
this region.
The signal region is the kinematic region in which data are compared to
the extrapolation from the control region to search for enhancements.
The search is divided into six overlapping regions of inclusive jet
multiplicity: $\NJ \ge 3$ to $\NJ \ge 8$.
The region with less than three jets is excluded because non-perturbative
gravitational states are unlikely to decay to one or two
jets~\cite{Giddings:2001bu} and the SM background would be larger with respect
to the signal than in the higher jet-multiplicity regions.
Other ATLAS searches~\cite{Aad:2014aqa} have set limits on this
low-multiplicity region.

The SM background in each signal region is estimated by fitting a function to
the data in the corresponding control region and then extrapolating the
resulting function to the signal region.
To fit the \HT{} distribution, a three-parameter $p_0, p_1, p_2$
empirical function

\begin{equation}\label{eq1}
\frac{\mathrm{d}N}{\mathrm{d}\HT} = \frac{p_0(1-x)^{p_1}}{x^{p_2}}\, ,
\end{equation}

\noindent
where $x \equiv \HT/\sqrt{s}$, is used.

The selection of the boundaries of the control region is based on 1)
stability of the extrapolation into the signal region of the function fit
to data with respect to small changes in the choice of control
region, and 2) minimisation of possible black hole (or string ball)
signal contamination in the control region.

The effect of possible signal contamination in the control region was
studied.
A string ball sample ($n=6$, $\Mth = 4.5\TeV$, $M_\mathrm{s} = 1.0\TeV$,
and $g_\mathrm{s} = 0.4$) was used in this study since it had the
largest fraction of events in the control region.
The number of string ball events was scaled down (by a factor of 27)
until its contribution to the signal region was at the current level of
detectability (three standard deviations above the expected background).
It was determined that this level of signal would not affect the fit and
its extrapolation by more than the statistical uncertainty if the upper
boundary on the control region is below $2.9\TeV$.
The value of the cross section used in this study has already been
ruled out at the 95\% confidence level~(CL)~\cite{Aad:2011bw,Aad:2012ic,Aad:2013lna,Aad:2014gka,Khachatryan:2010wx,Chatrchyan:2012taa,Chatrchyan:2013xva}.

\begin{figure}[htb]
\begin{center}
\includegraphics[width=0.43\textwidth]{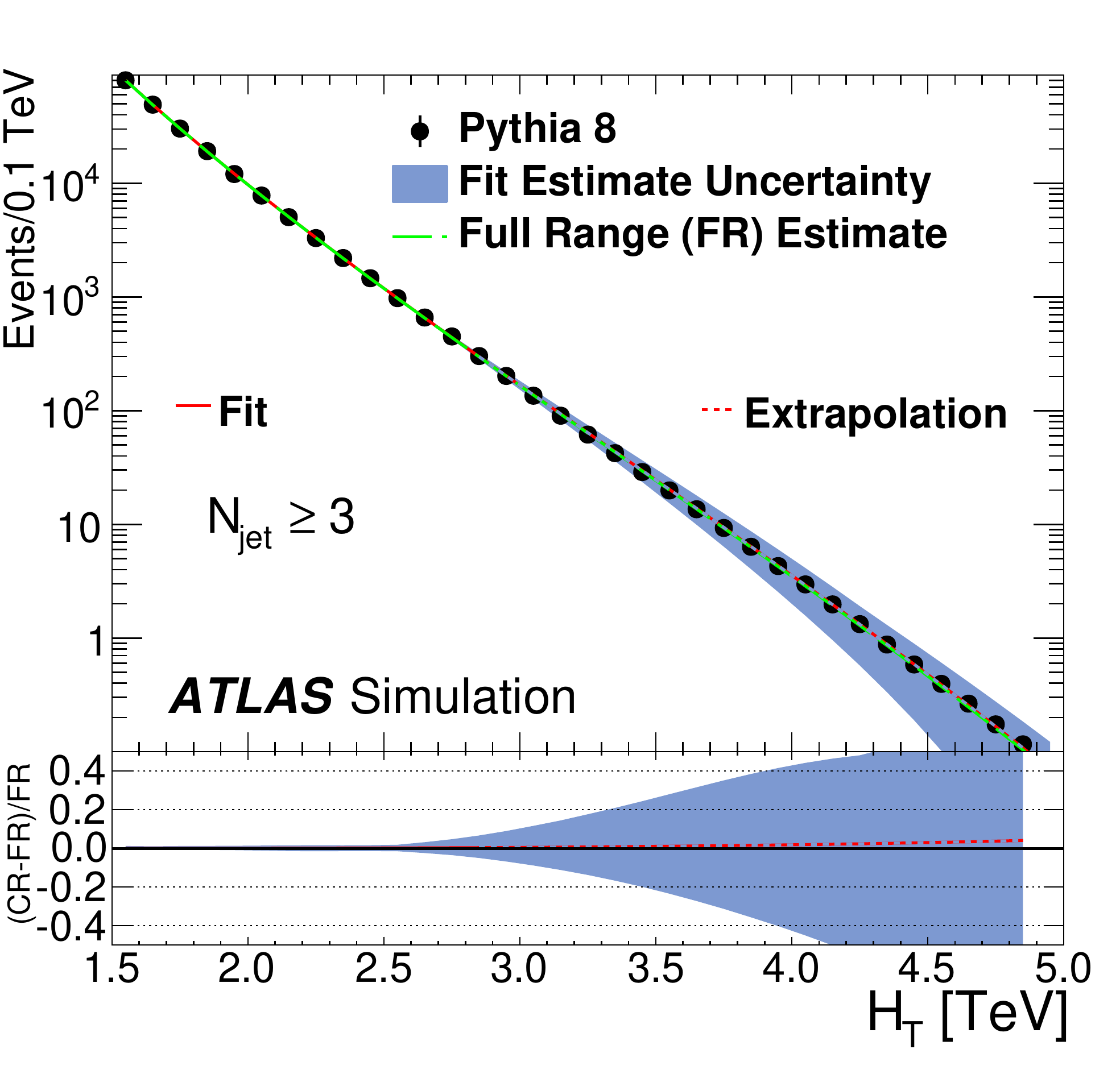}
\includegraphics[width=0.43\textwidth]{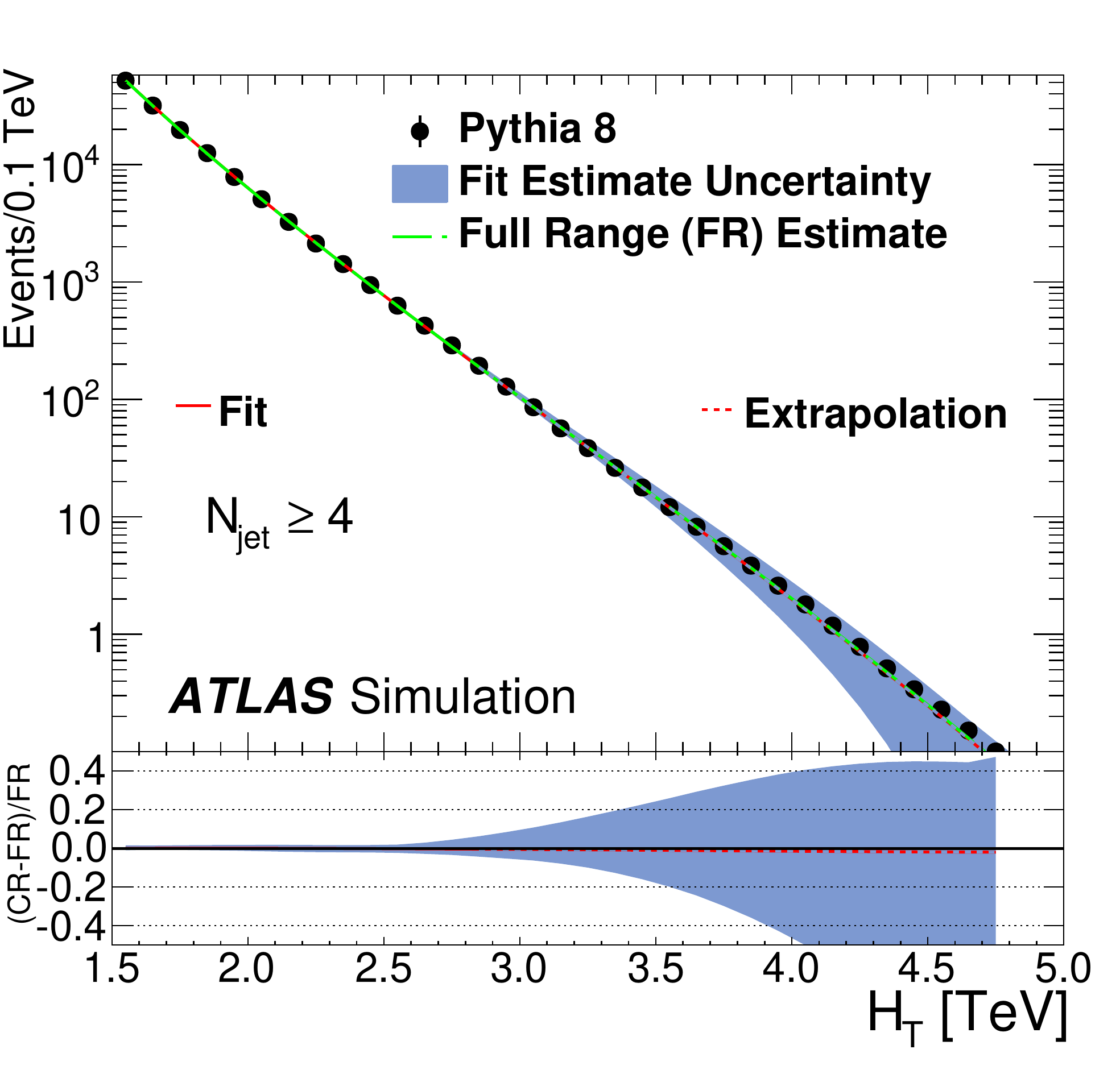}
\includegraphics[width=0.43\textwidth]{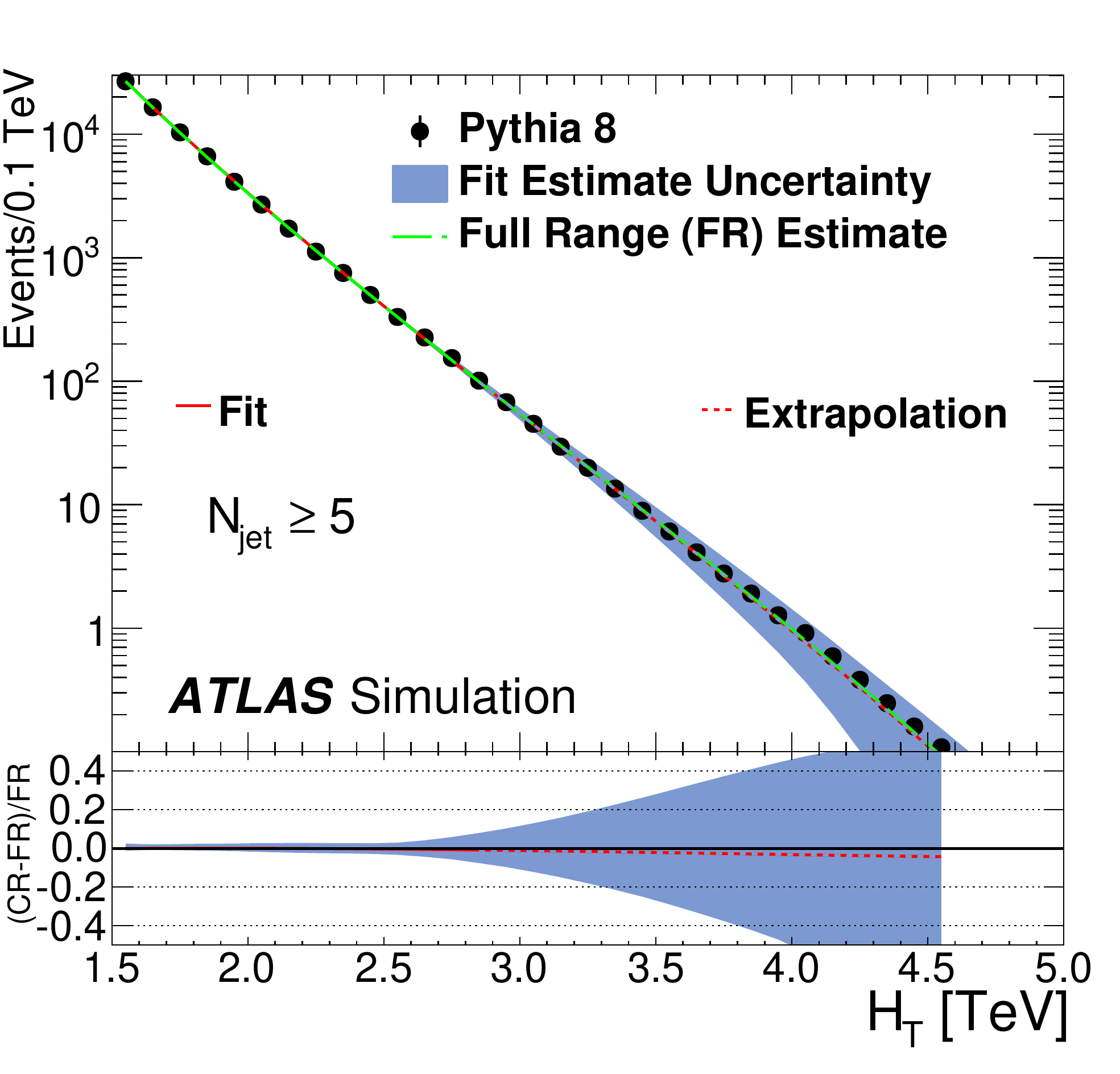}
\includegraphics[width=0.43\textwidth]{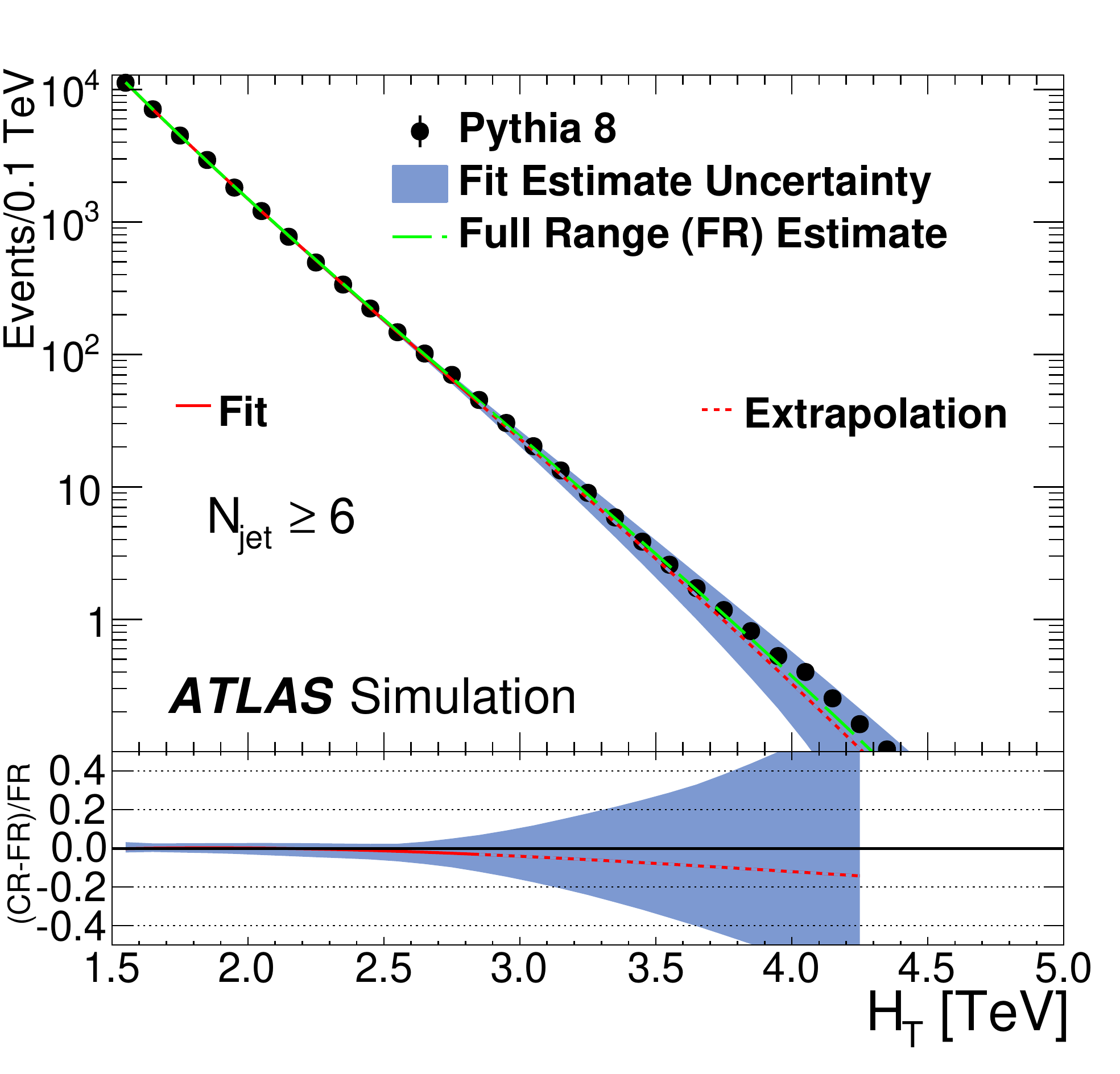}
\includegraphics[width=0.43\textwidth]{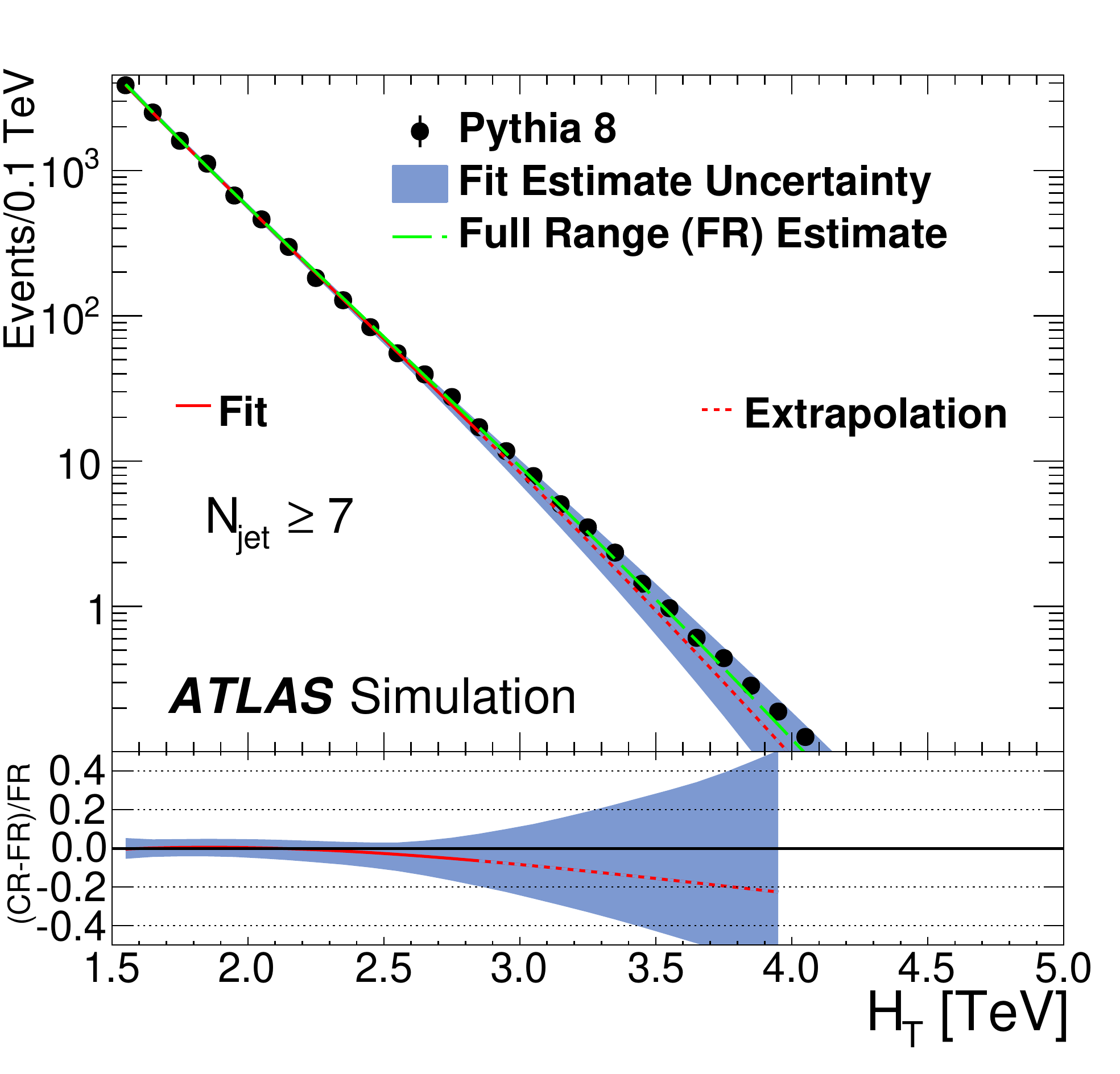}
\includegraphics[width=0.43\textwidth]{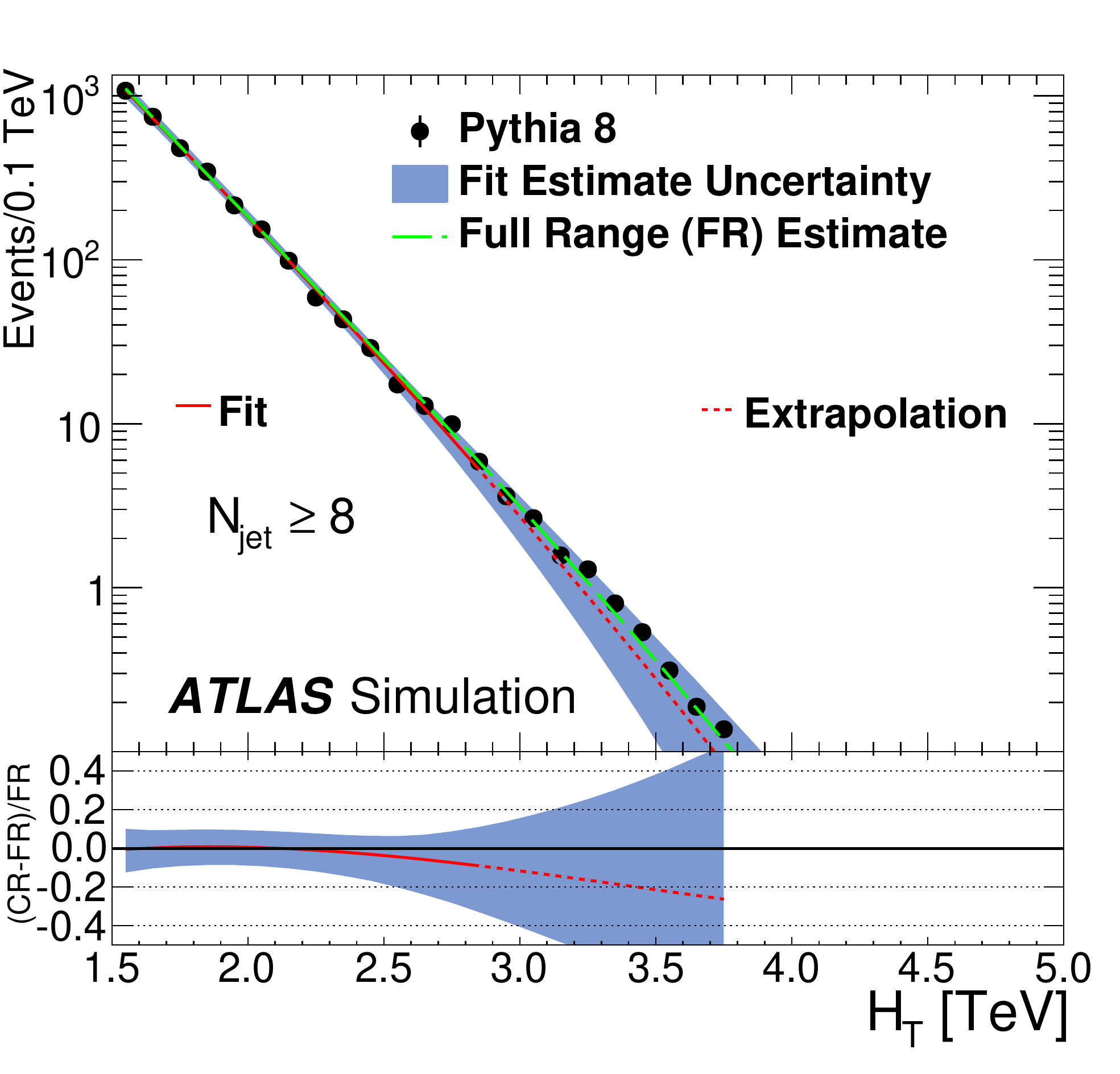}
\end{center}
\caption{The \HT{} distributions, showing a comparison of the full range
(FR) fit from $\HT=\HTmin\TeV$ to the last predicted data value with
the extrapolation of the fit from the control region (CR) $\HTmin < \HT
< \HTmax\TeV$ into the signal region ($\HT > 3.0\TeV$) for \pythia~8
simulated events.
The uncertainty band includes all the uncertainties described in
section~\ref{sec:sys}.}
\label{fig2}
\end{figure}

Although the background estimate only relies on data, the validity of the
assumption that the fit in the control region can be used to estimate the
background in the signal region was tested using \pythia~8,
\herwigpp, and \alpgen MC simulated events.
Since the multi-jet SM simulated events do not contain a signal, the
results of the fit to the entire \HT{} distribution can be compared to
the results of the fit to only the control region and extrapolating into
the signal region.
As an example, the results for \pythia~8 dijet simulated events is shown
in figure~\ref{fig2}.
These studies show that the fit extrapolation approximates the MC
simulated events in the signal region to within 20\%.
This difference is covered by the uncertainties, which are described in
the next section.

\section{Systematic uncertainties}
\label{sec:sys}

In addition to the statistical uncertainties from the limited number of
events in the control region, systematic uncertainties arising from the
choice of control region and the choice of fit function are considered.
To estimate the effect of limited number of events in the \HT{}
distributions on the fit, pseudo \HT-distributions with the same number
of events as data are generated using the fit to data in the control
region as a probability density function.
Each pseudo \HT-distribution is fit, and the predicted number of events
in each \HT{} bin is calculated and subtracted from the prediction from
the fit to data.
A distribution of these differences between the fit to data and the fit
to pseudo distributions is used to derive an uncertainty on the fit in
each \HT{} bin.
The value of the deviation corresponding to 68\% of the area, about the nominal fit prediction, under the
distribution is taken as the asymmetric statistical uncertainty on the fit.
In the signal regions, the statistical uncertainty on the fit rises
from 5\% at the lower edge of the \HT{} range to 17\% at the limit of no
data for $\NJ\ge 3$, and from 37\% to 67\% for $\NJ\ge 8$.

A systematic uncertainty is assigned due to the \HT{} range chosen for
the control regions.
To estimate the uncertainty on the nominal choice, the data are fit in
all eight possible \HT{} control regions by increasing and decreasing
the fit range by $0.1\TeV$, and shifting it by $\pm 0.1\TeV$.
Each fit is used to predict the number of events in each \HT{} bin.
The control region predicting the largest number of events and the
control region predicting the smallest number of events for each \HT{}
bin provide an estimate of the asymmetric uncertainty due to the choice
of control region.
In the signal regions, the systematic uncertainty due to the choice of
control region rises from 2\% at the lower edge of the \HT{} range to 8\%
at the limit of no data for $\NJ\ge 3$, and from 28\% to 53\% for
$\NJ\ge 8$.

To estimate the uncertainty in the analysis due to the choice of fit
function, alternative fit functions are considered.
Alternative fit functions are chosen such that in multi-jet simulated
events (\pythia~8 and \herwigpp) they provide a good fit in the control
region and the extrapolation of the fit provides a good description of
the simulated events in the corresponding signal region.
In addition, the function should also provide a good description of the
data in the control region.
Only functional forms that fulfil these criteria for at least one
inclusive jet multiplicity region are considered, and these are:

\begin{equation}\label{eq5}
p_0(1-x)^{p_1} e^{p_2x^2}\, ,
\end{equation}

\begin{equation}\label{eq11}
p_0(1-x)^{p_1}x^{p_2x}\, ,
\end{equation}

\begin{equation}\label{eq12}
p_0 (1-x)^{p_1} x^{p_2\ln x}\, ,
\end{equation}

\begin{equation}\label{eq15}
p_0(1-x)^{p_1}(1+x)^{p_2x}\, ,
\end{equation}

\begin{equation}\label{eq16}
p_0(1-x)^{p_1}(1+x)^{p_2\ln x}\, ,
\end{equation}

\begin{equation}\label{eq17}
\frac{p_0}{x}(1-x)^{[p_1-p_2\ln x]}\, ,
\end{equation}

\begin{equation}\label{eq18}
\frac{p_0}{x^2}(1-x)^{[p_1-p_2\ln x]}\, .
\end{equation}

\begin{figure}[htb]
\begin{center}
\includegraphics[width=0.44\textwidth]{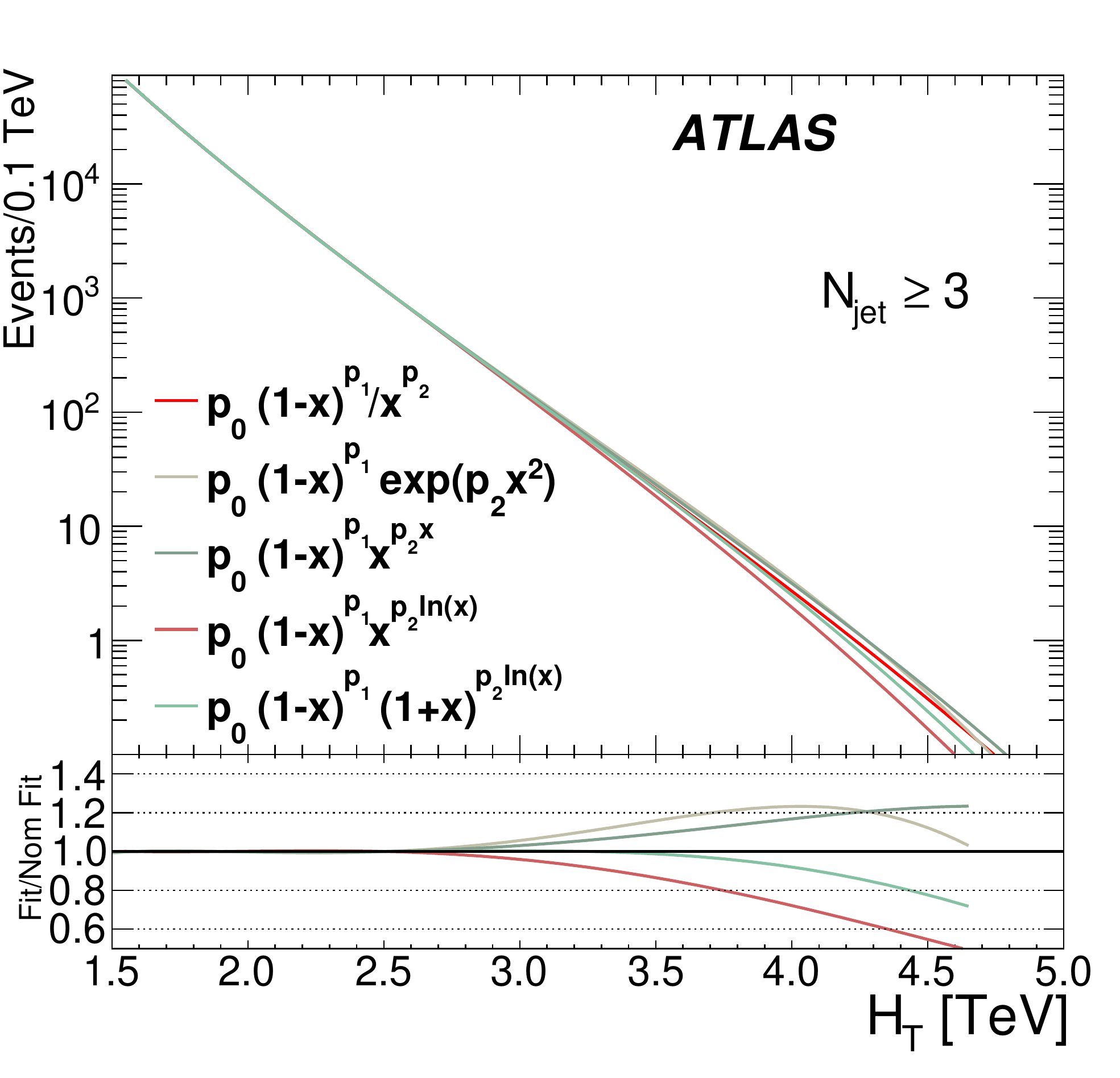}
\includegraphics[width=0.44\textwidth]{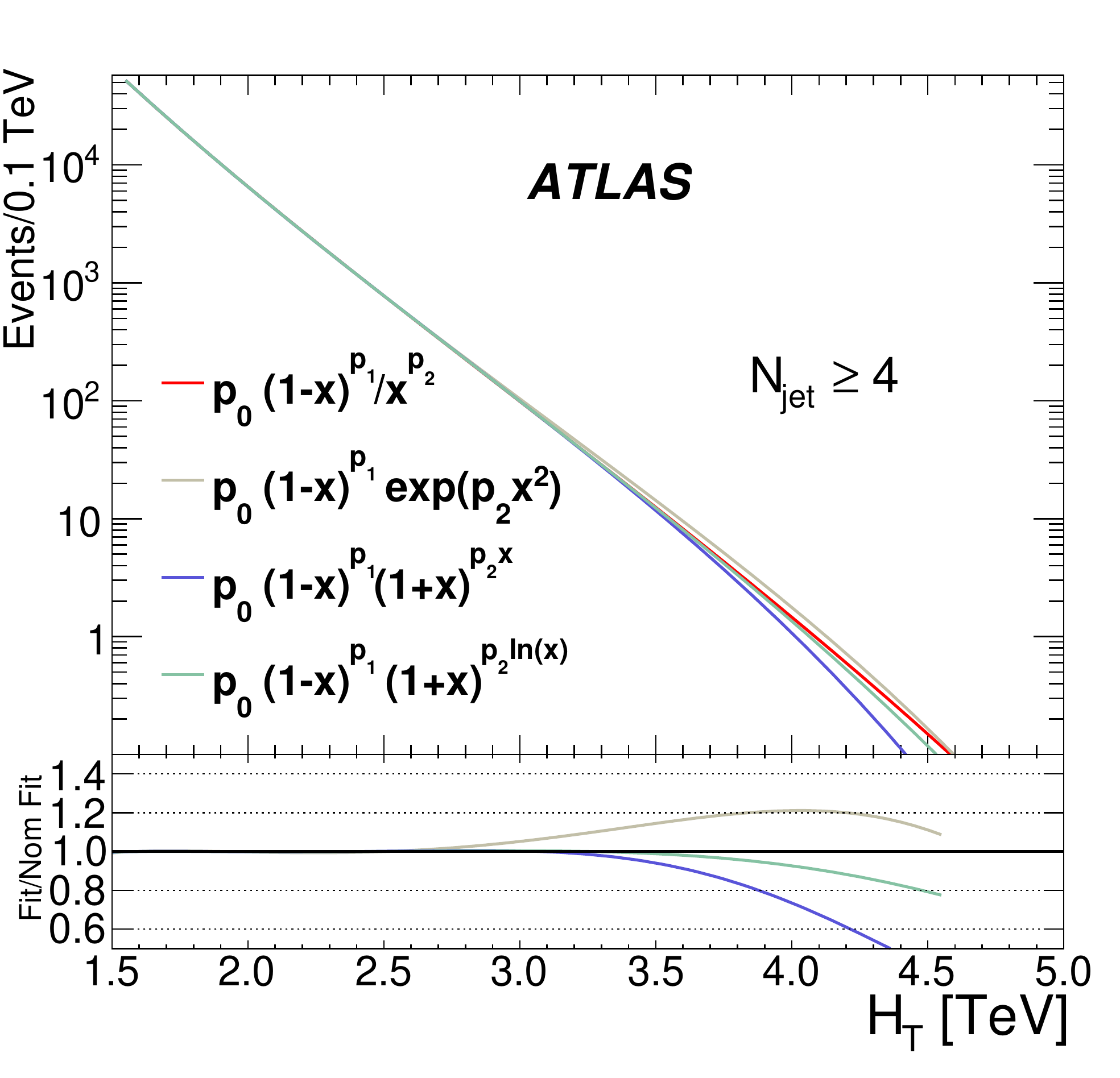}
\includegraphics[width=0.44\textwidth]{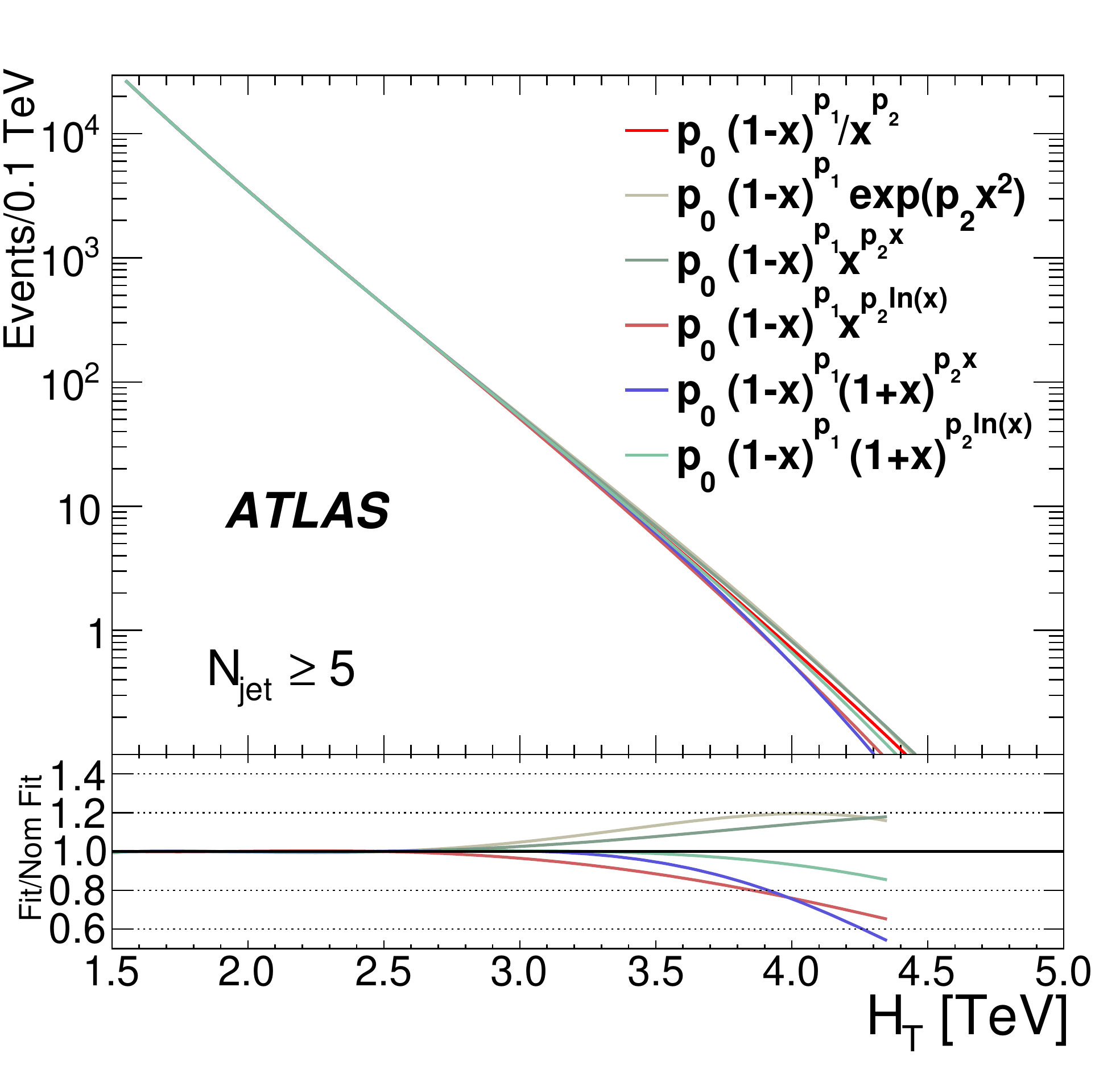}
\includegraphics[width=0.44\textwidth]{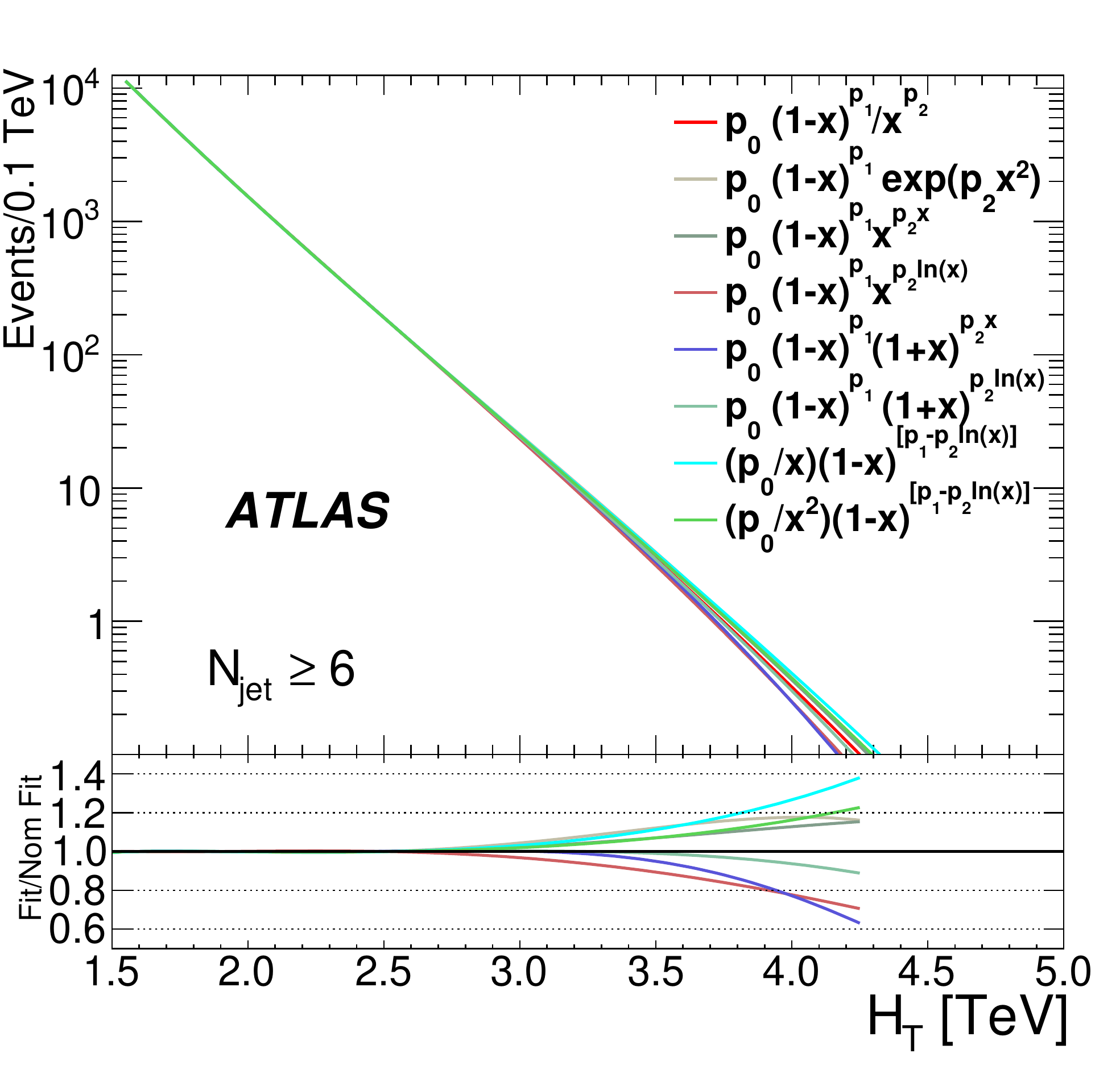}
\includegraphics[width=0.44\textwidth]{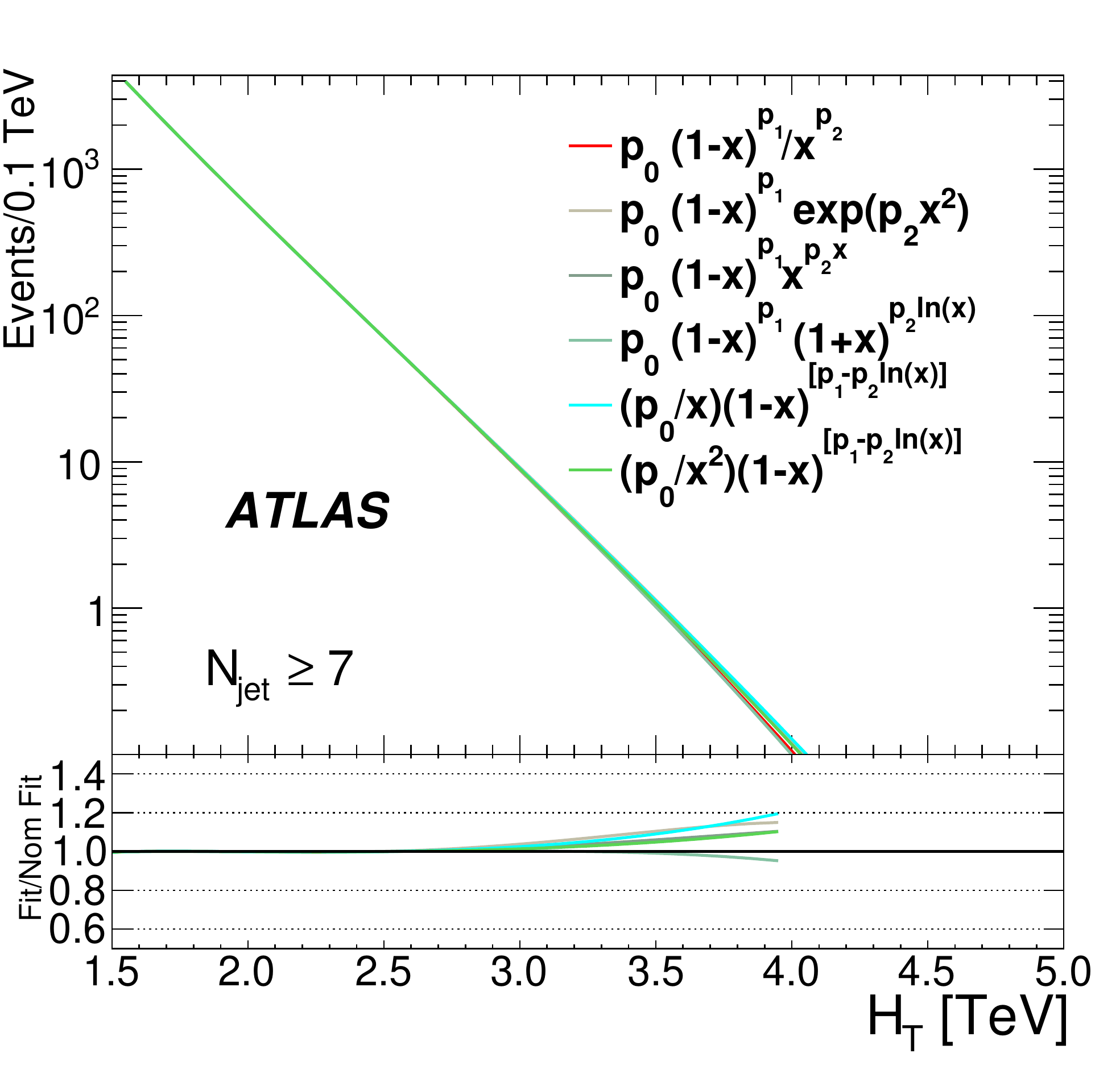}
\includegraphics[width=0.44\textwidth]{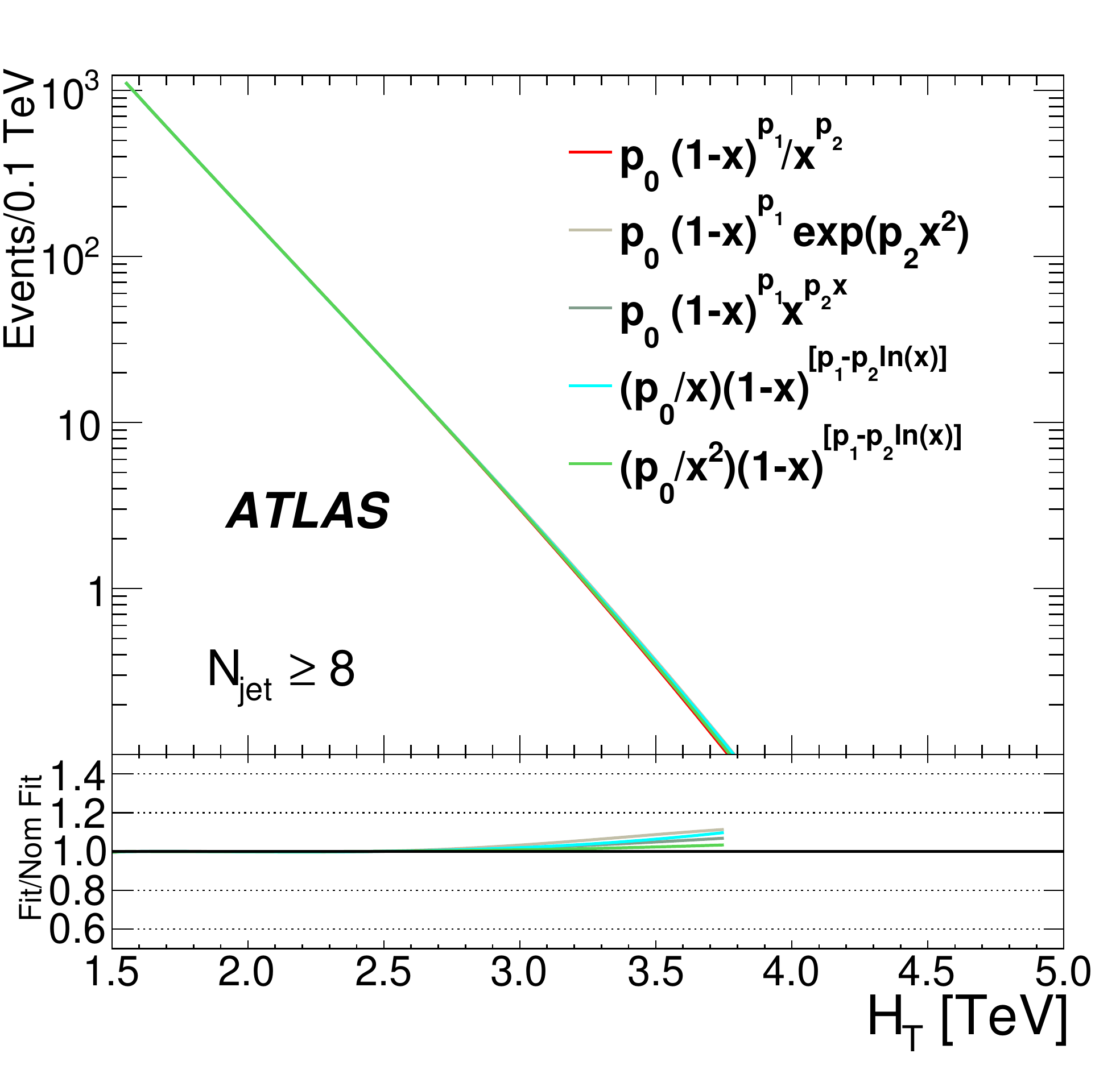}
\end{center}
\caption{The \HT{} distributions, showing alternative fit functions for
different inclusive jet multiplicities \NJ.
The systematic uncertainty due to the choice of fit function versus exclusive
\HT{} is given by the envelope of all the functions.}
\label{fig3}
\end{figure}

\noindent
The systematic uncertainty due to the choice of fit function is assigned
as the envelope of all alternative fit functions around the nominal
function when fit to data in the control region.
Figure~\ref{fig3} shows the alternative fit functions for different
inclusive jet multiplicities.
In the signal regions, the systematic uncertainty due to the choice of fit
function rises from 10\% at the lower edge of the \HT{} range to 46\% at
the limit of no data for $\NJ\ge 3$, and from 6\% to 10\% for $\NJ\ge 8$.

The jet energy scale (JES) uncertainties and jet energy resolution
(JER) uncertainty are used in the MC validation methods and applied
to the MC signal samples for the model-dependent limit calculations only.
In the signal regions, the JES uncertainty rises from 1\% at the lower
edge of the \HT{} range to 22\% at the limit of no data, and the JER
uncertainty rises from 0.6\% to 2\% for $\NJ\ge 3$.
For $\NJ \ge 8$, the JES uncertainty rises from 21\% at the lower edge
of the \HT{} range to 34\% at the limit of no data, and the JER
uncertainty rises from 7\% to 11\%.

\section{Results}
\label{sec:result}

\begin{figure}[htb]
\begin{center}
\includegraphics[width=0.40\textwidth]{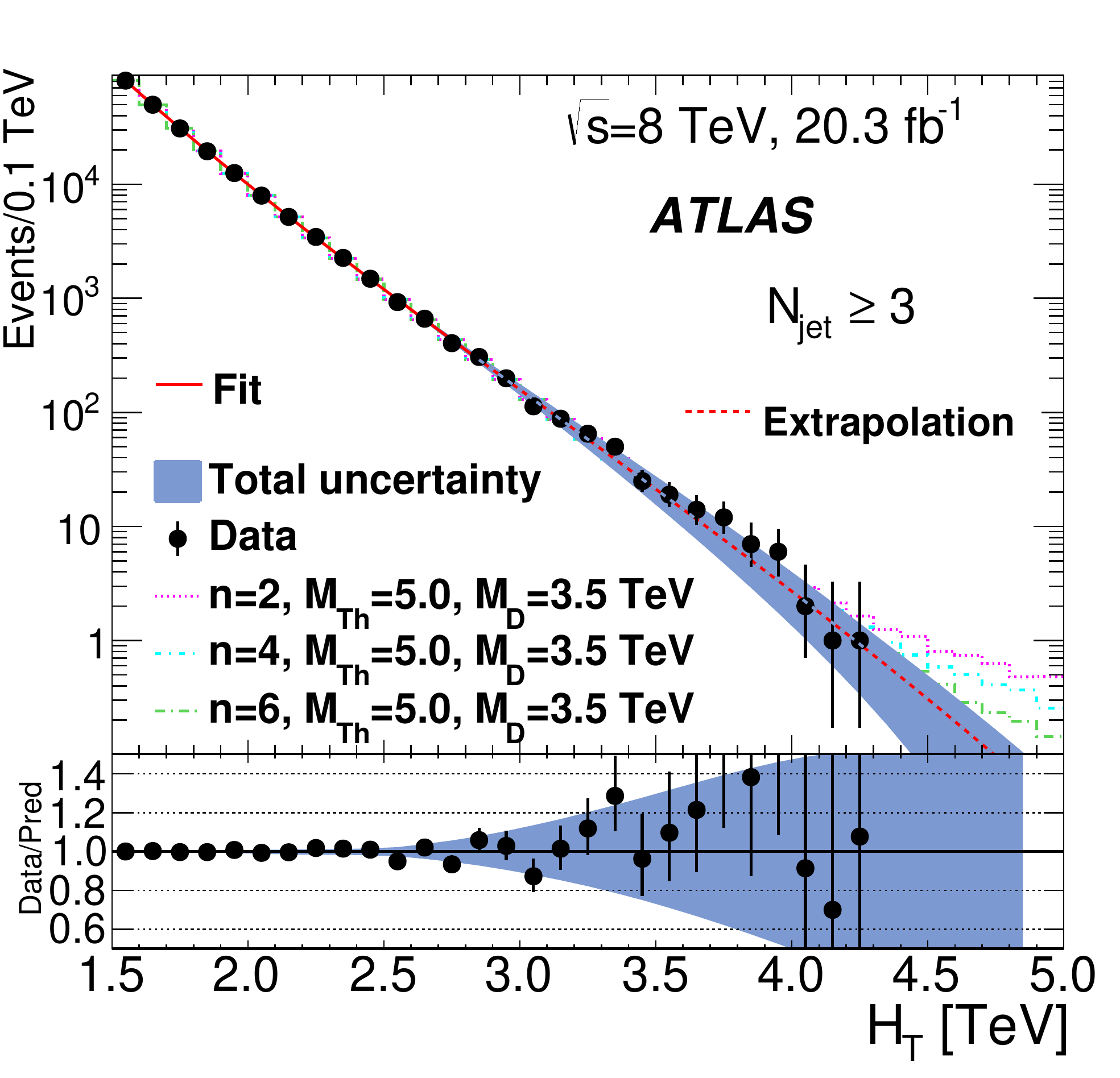}
\includegraphics[width=0.40\textwidth]{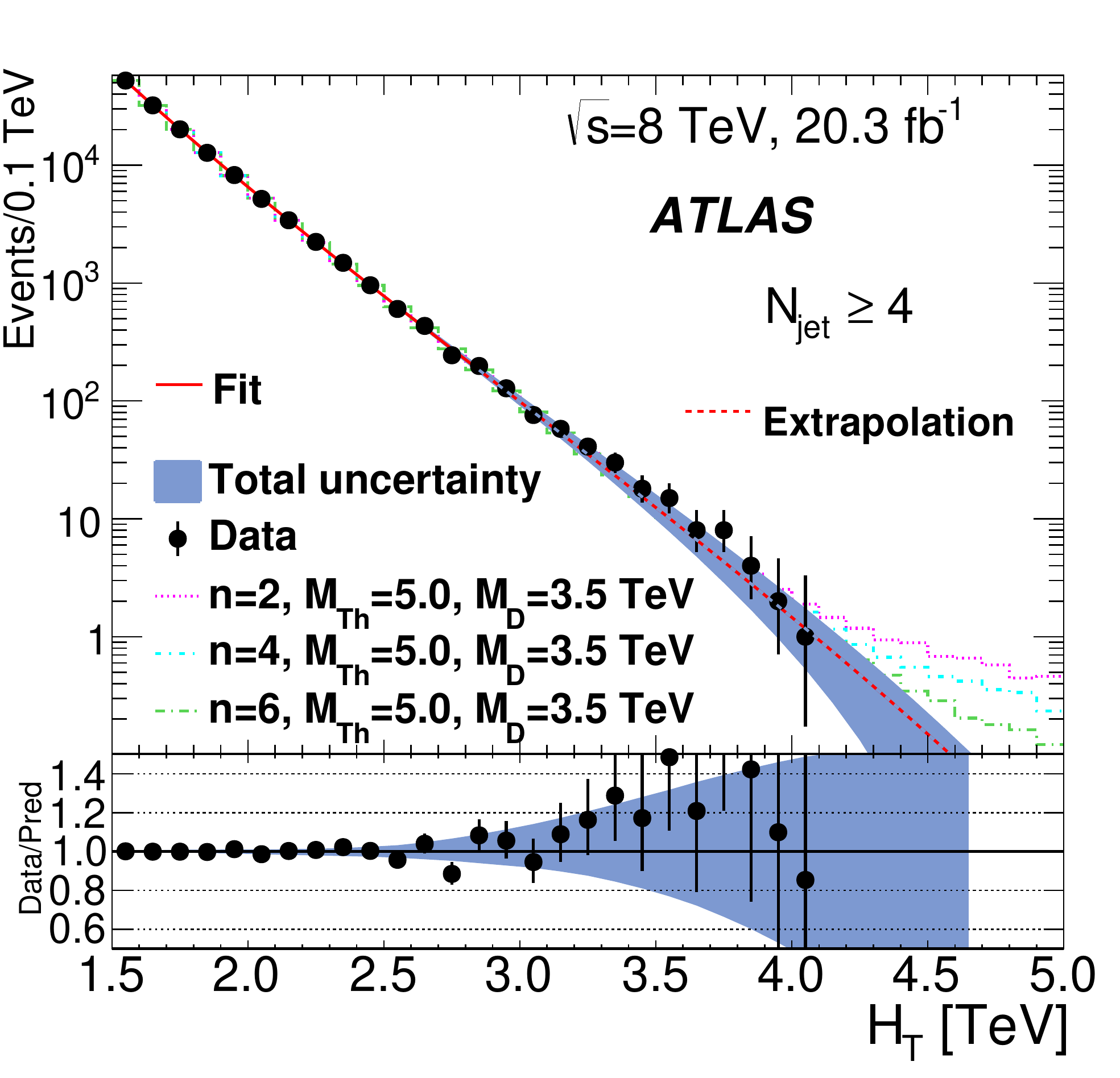}
\includegraphics[width=0.40\textwidth]{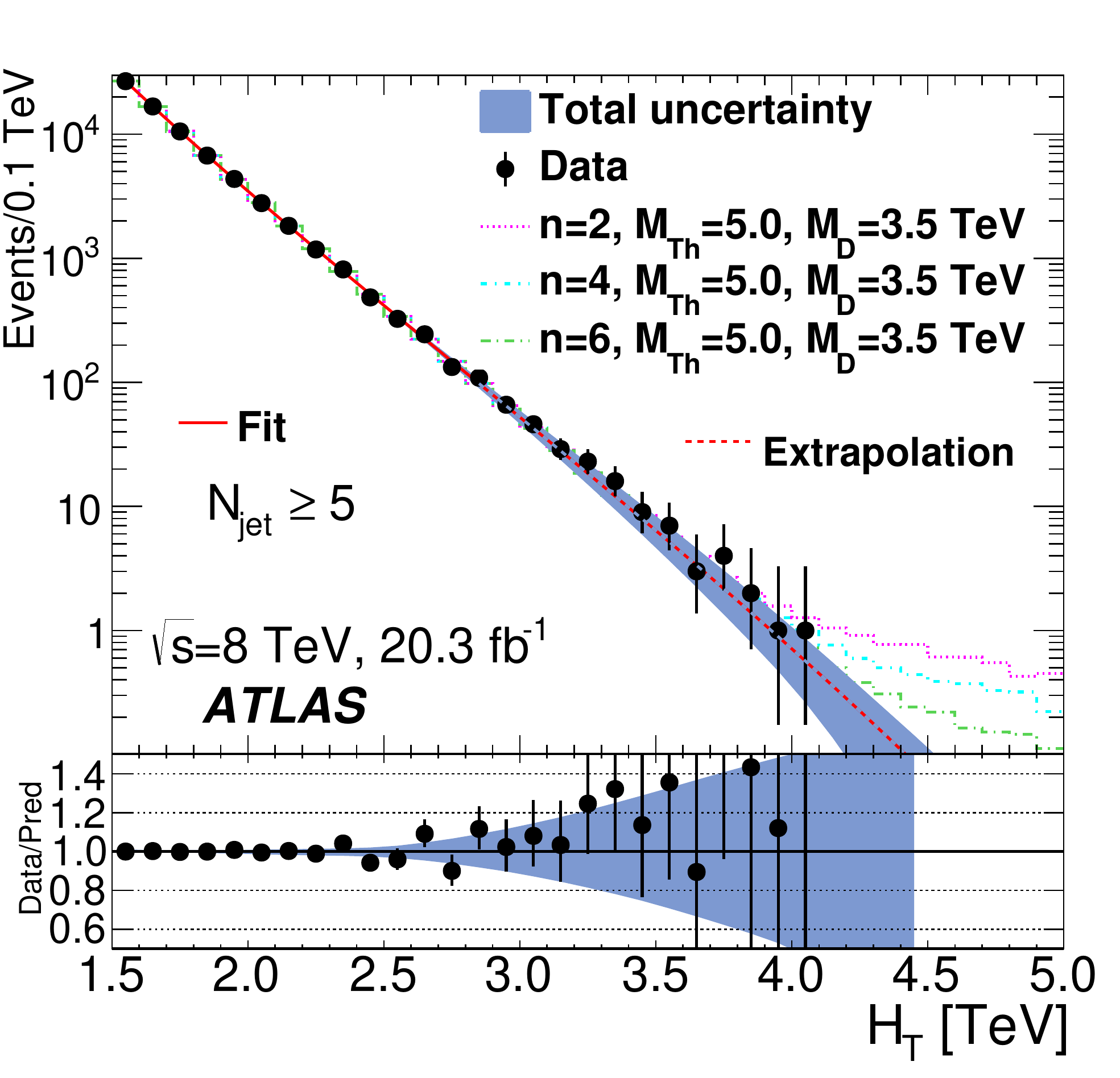}
\includegraphics[width=0.40\textwidth]{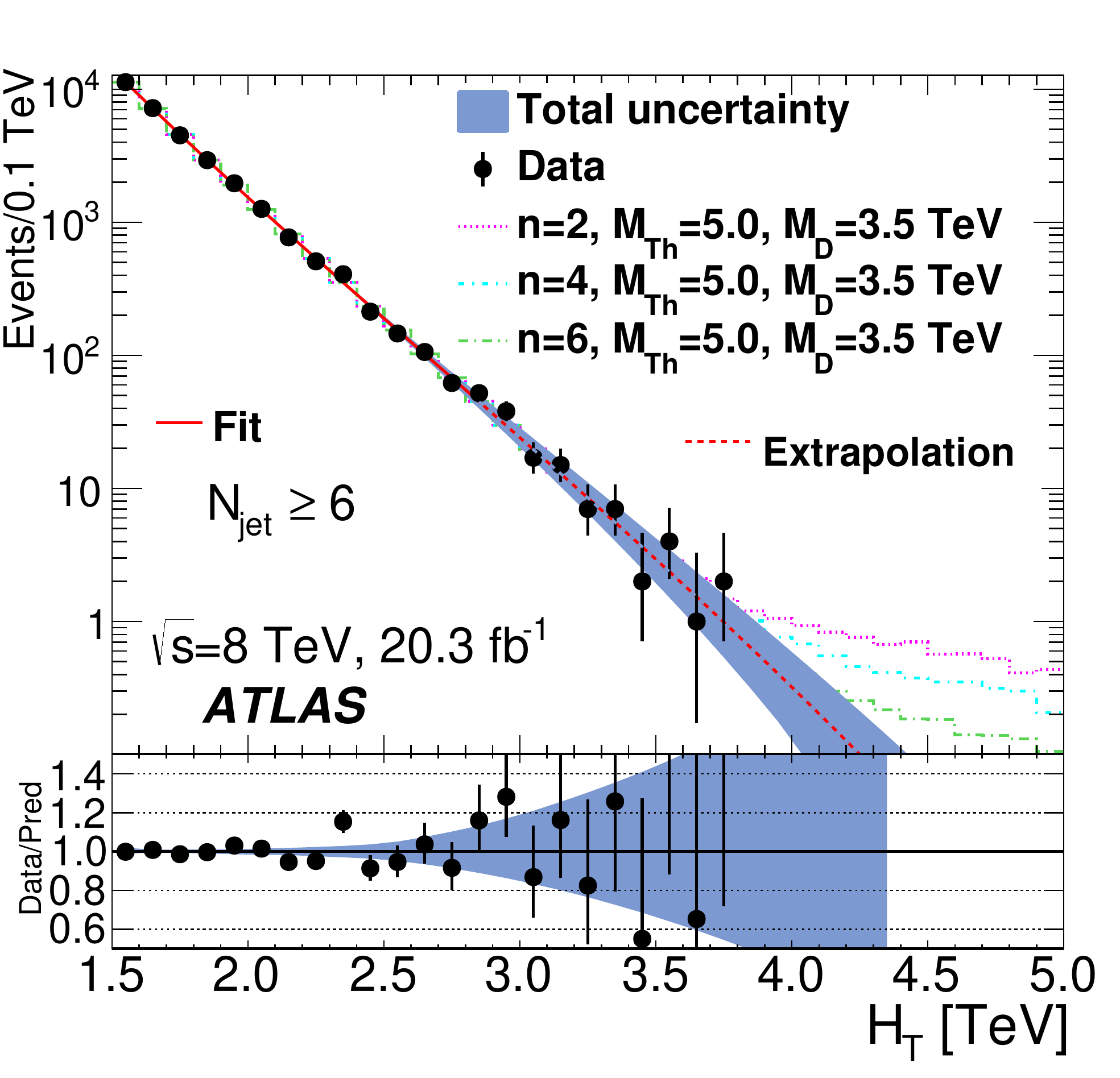}
\includegraphics[width=0.40\textwidth]{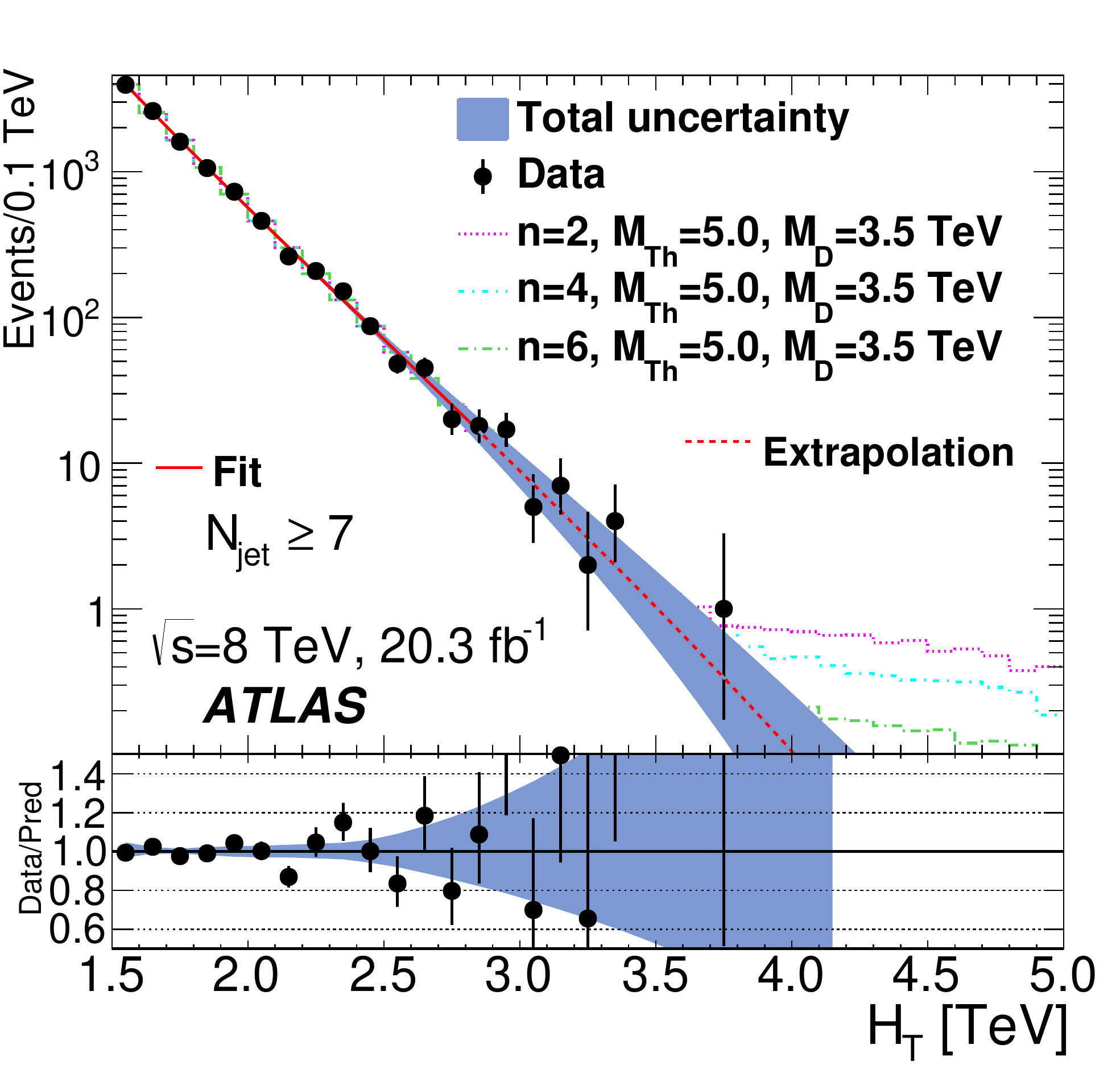}
\includegraphics[width=0.40\textwidth]{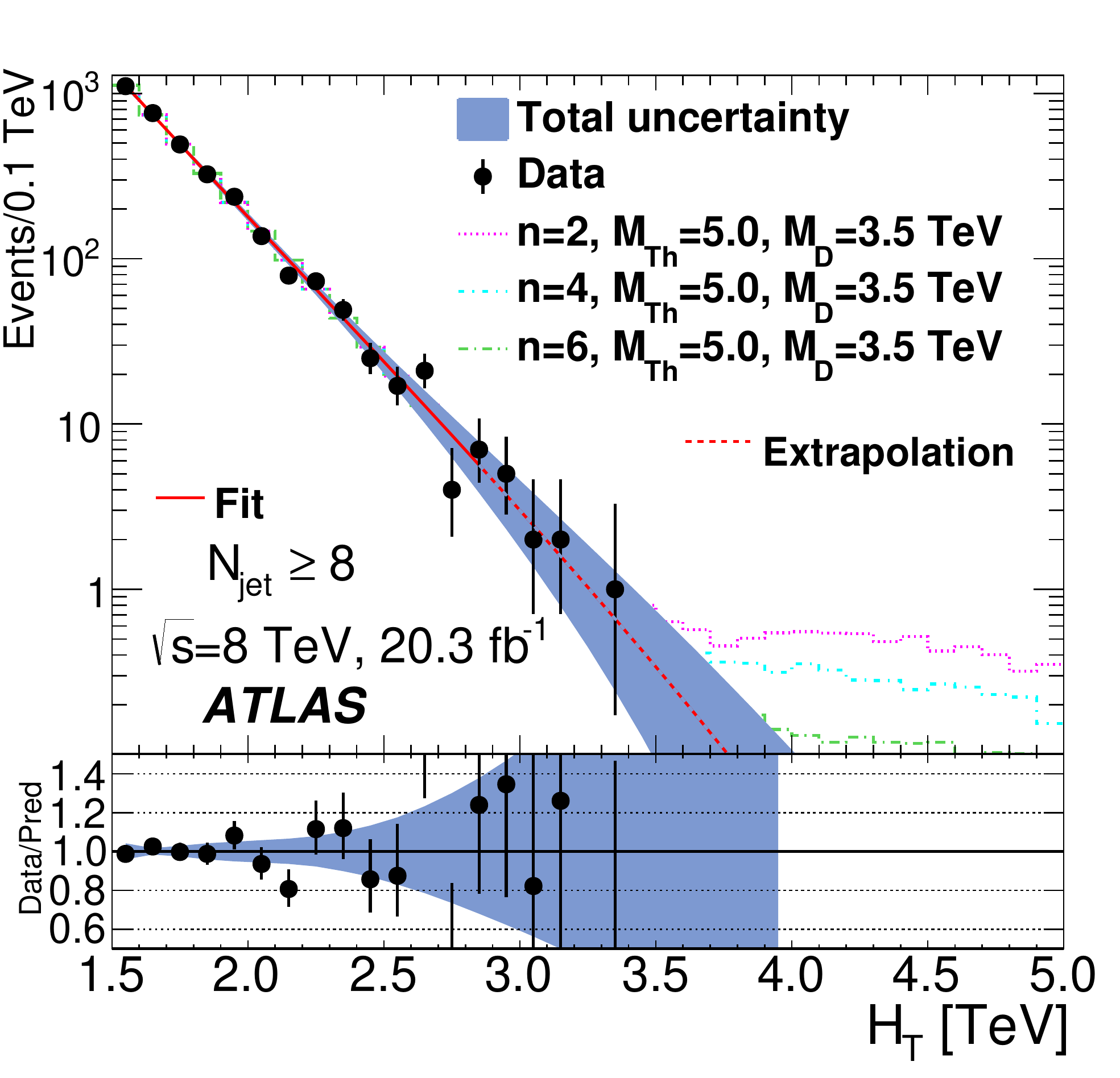}
\caption{The \HT{} distributions, showing the data and extrapolated fits
from the control region $\HTmin < \HT < \HTmax\TeV$ into the signal
region $\HT > 3.0$ for each inclusive jet multiplicity \NJ.
The uncertainty on the data points in both the distribution and ratio
are due to the statistical uncertainty on the data only.
The uncertainty band includes all uncertainties on the background prediction.
Also shown are the expected black hole signals for three parameter sets of the
\charybdis non-rotating black hole model.}
\label{fig4}
\end{center}
\end{figure}

Figure~\ref{fig4} shows the extrapolation of the fits in the control region
to the signal region with all uncertainties included.
No data events are observed above $\HT = 4.3\TeV$, in agreement with the
background estimate.

To test the consistency of the data with the null-hypothesis
(background-only hypothesis) a hyper-test statistic $t =
-\ln[p\text{-}\mathrm{value}^\mathrm{min}]$ is defined where
$p$-value$^\mathrm{min}$ is the minimum local $p$-value in any inclusive \NJ{}
and \HT{} region~\cite{2011arXiv1101.0390C}.
The \HT regions can range from a single $0.1\TeV$ bin to the entire range
containing data.
The hyper-test statistic takes account of the trials factor
(look-elsewhere effect).
The most significant discrepancy in the observed signal region
distributions is an  excess in the interval $3.2\TeV$ to $3.9\TeV$,
for $\NJ\ge 4$.
This enhancement corresponds to a local $p$-value of 0.0043 which
corresponds to a significance of 2.6 standard
deviations~\cite{2012EPJP..127...25C} compared to the most probable value
for the null-hypothesis of $2.3\sigma$.
The corresponding $t$-value for the data is 5.4, and for the
null-hypothesis the probability to find a value equal or greater than
5.4 is 0.4.
This test shows that no significant excess is observed beyond the SM
expectations for all choices of \HT{} and inclusive $\NJ$ signal
regions.

Since black holes or string balls are likely to appear as an enhancement
in the tail of \HT{} distributions, rather than as resonances,
production upper limits are set in bins of inclusive \HT{} (\HTinc{}),
rather than \HT.
The predicted number of events in each \HTinc{} bin in the signal region
is obtained by integrating the fit function from the \HT{} bin of
interest up to the kinematic limit of $8\TeV$.
The same integral is performed for the maximum and minimum number of
predicted events obtained from the statistical uncertainty and each
systematic uncertainty.
The differences between the maximum (or minimum) number of predicted events
and the nominal number of predicted events for each uncertainty are
added linearly to obtain the total uncertainty on the number of
predicted events in each \HTinc{} bin.
The uncertainties are added linearly since each is obtained from the same
data in the control region.

\begin{figure}[htb]
\begin{center}
\includegraphics[width=0.41\textwidth]{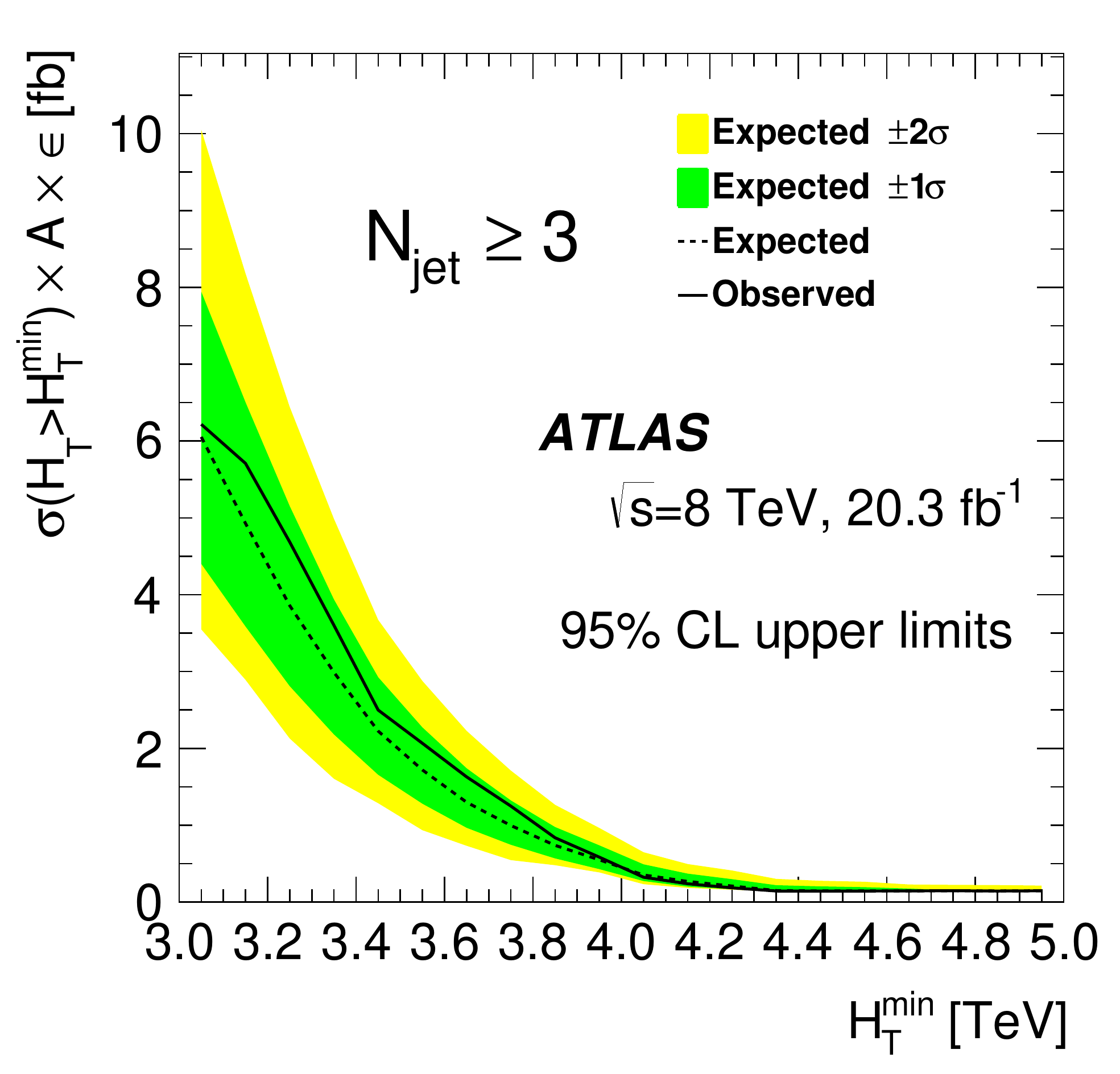}
\includegraphics[width=0.41\textwidth]{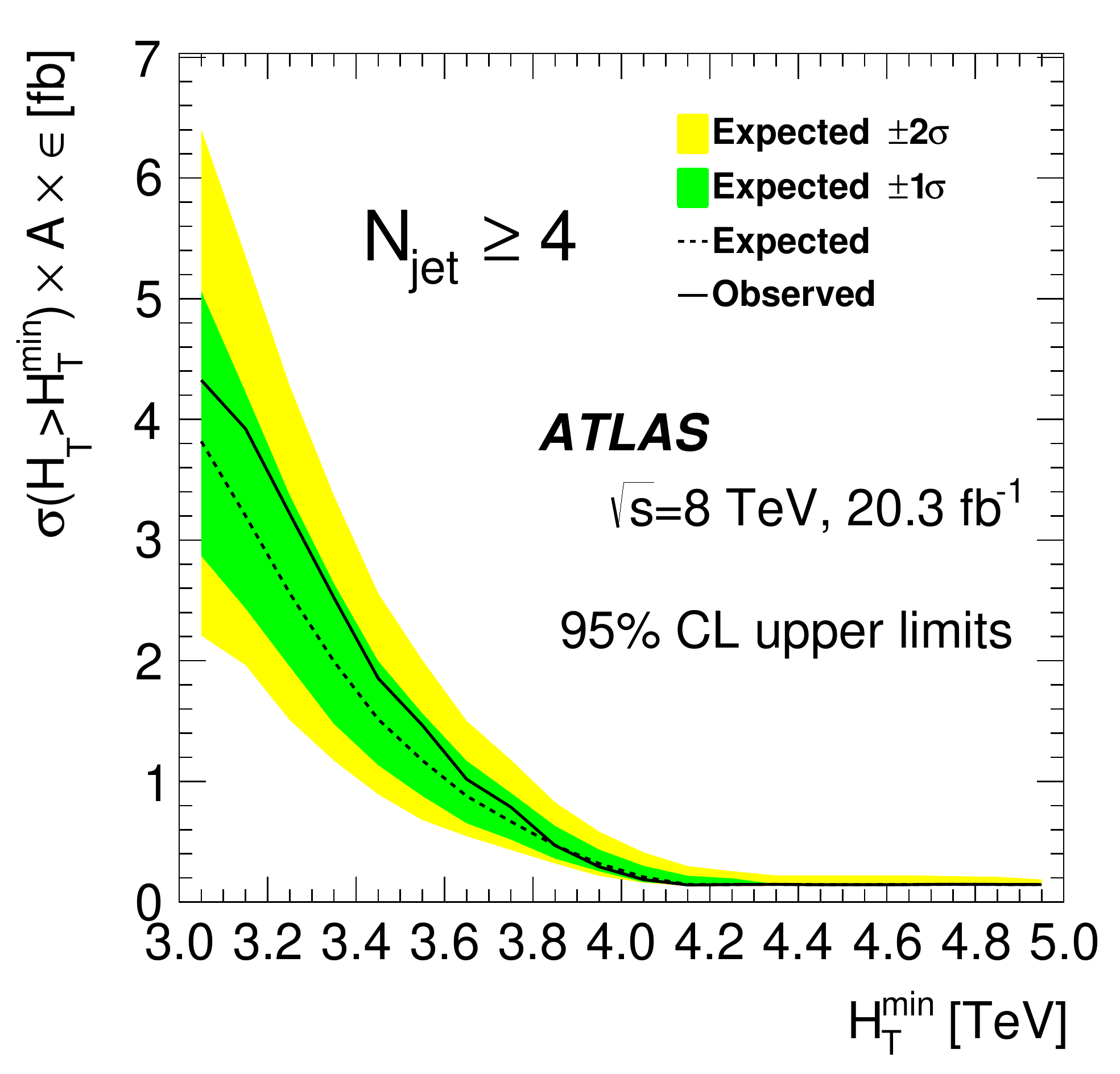}
\includegraphics[width=0.41\textwidth]{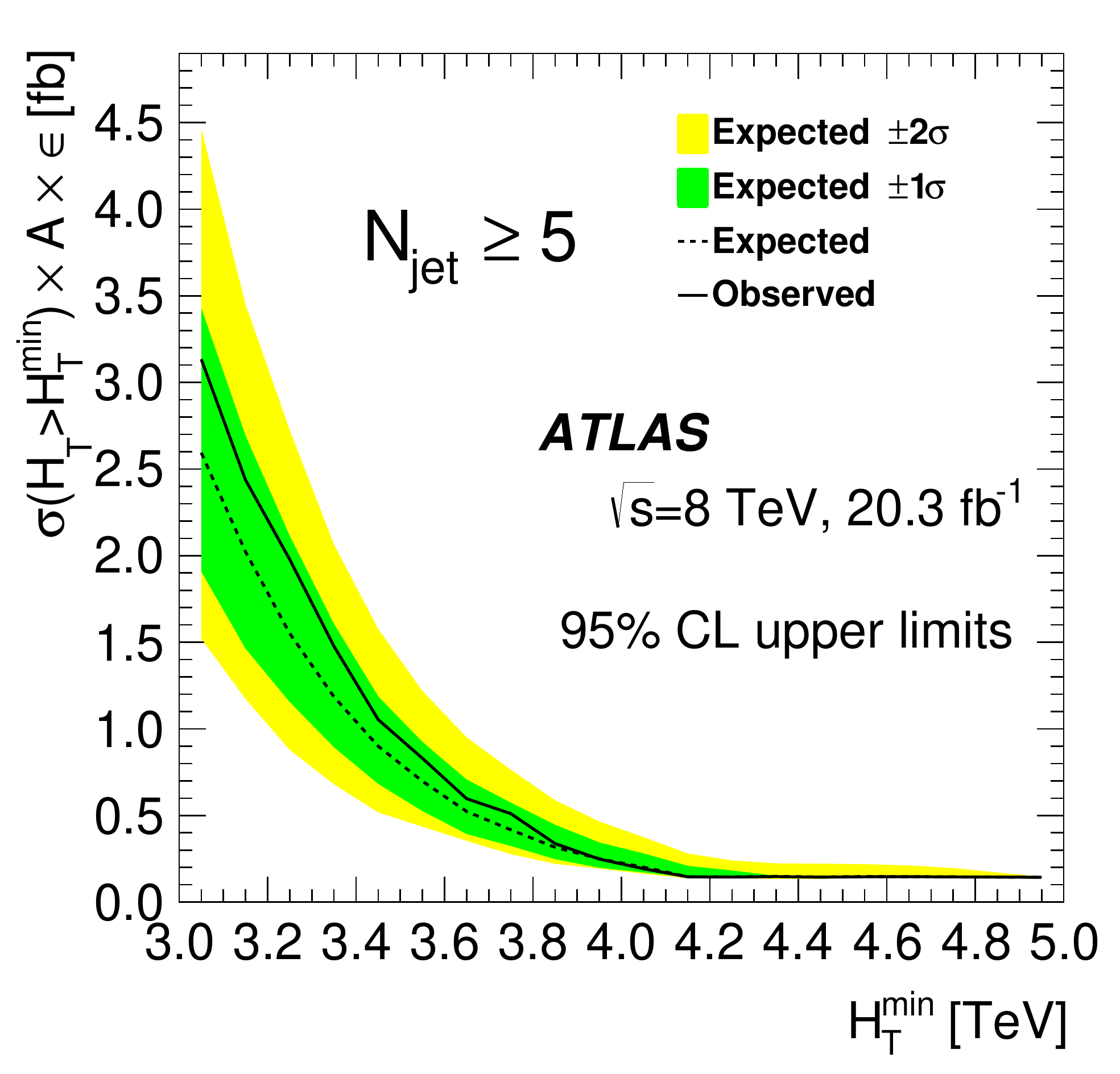}
\includegraphics[width=0.41\textwidth]{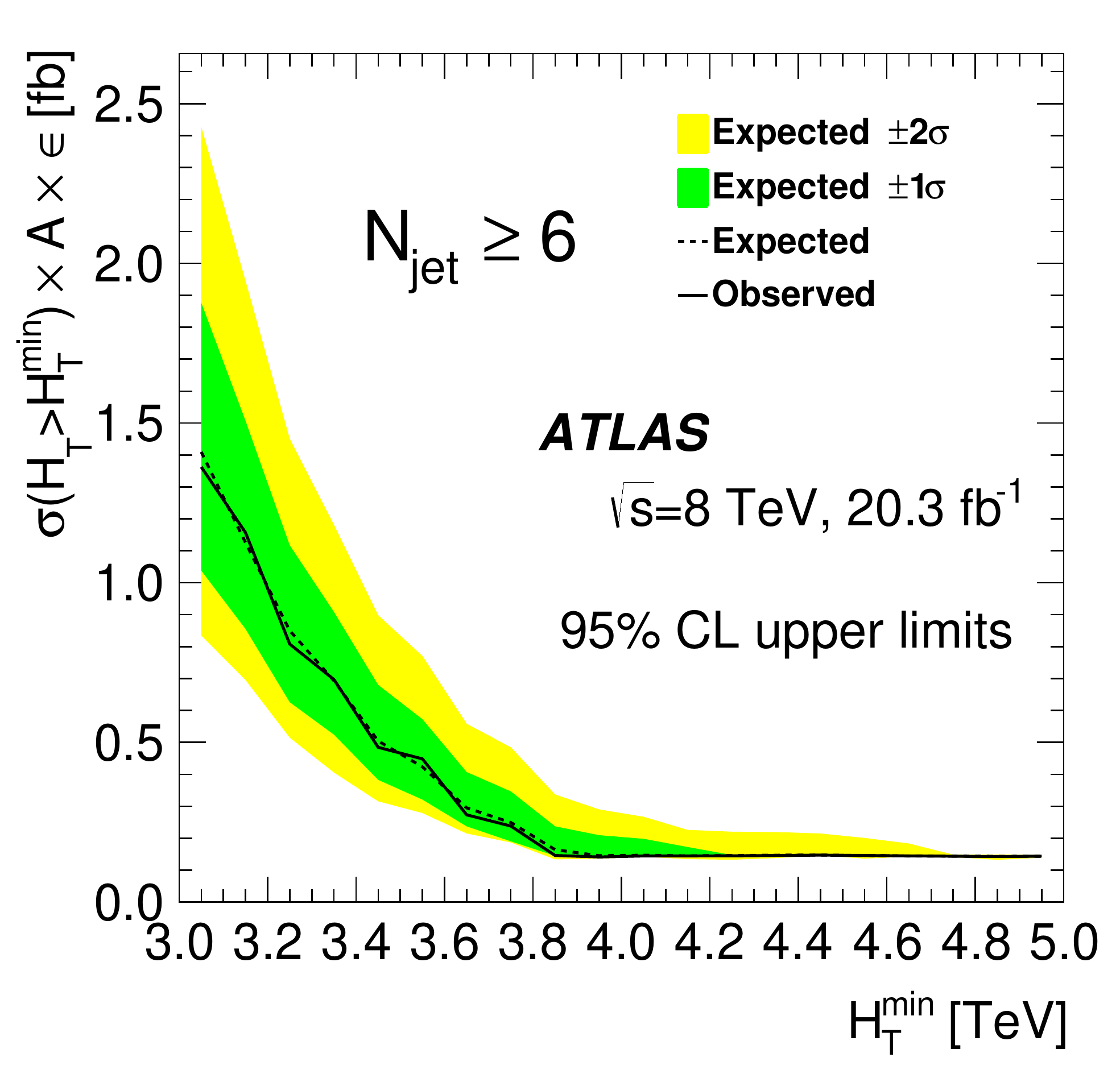}
\includegraphics[width=0.41\textwidth]{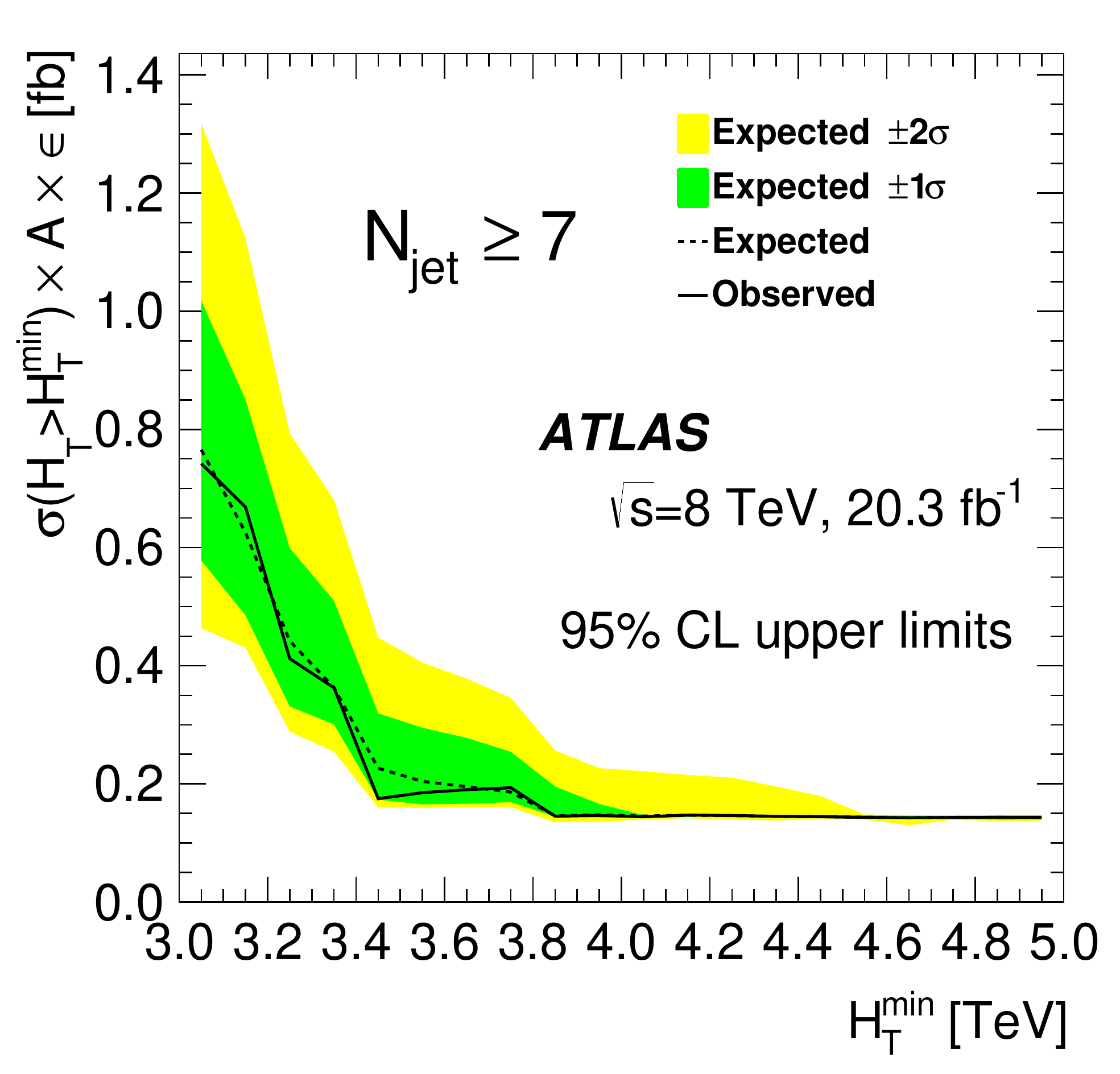}
\includegraphics[width=0.41\textwidth]{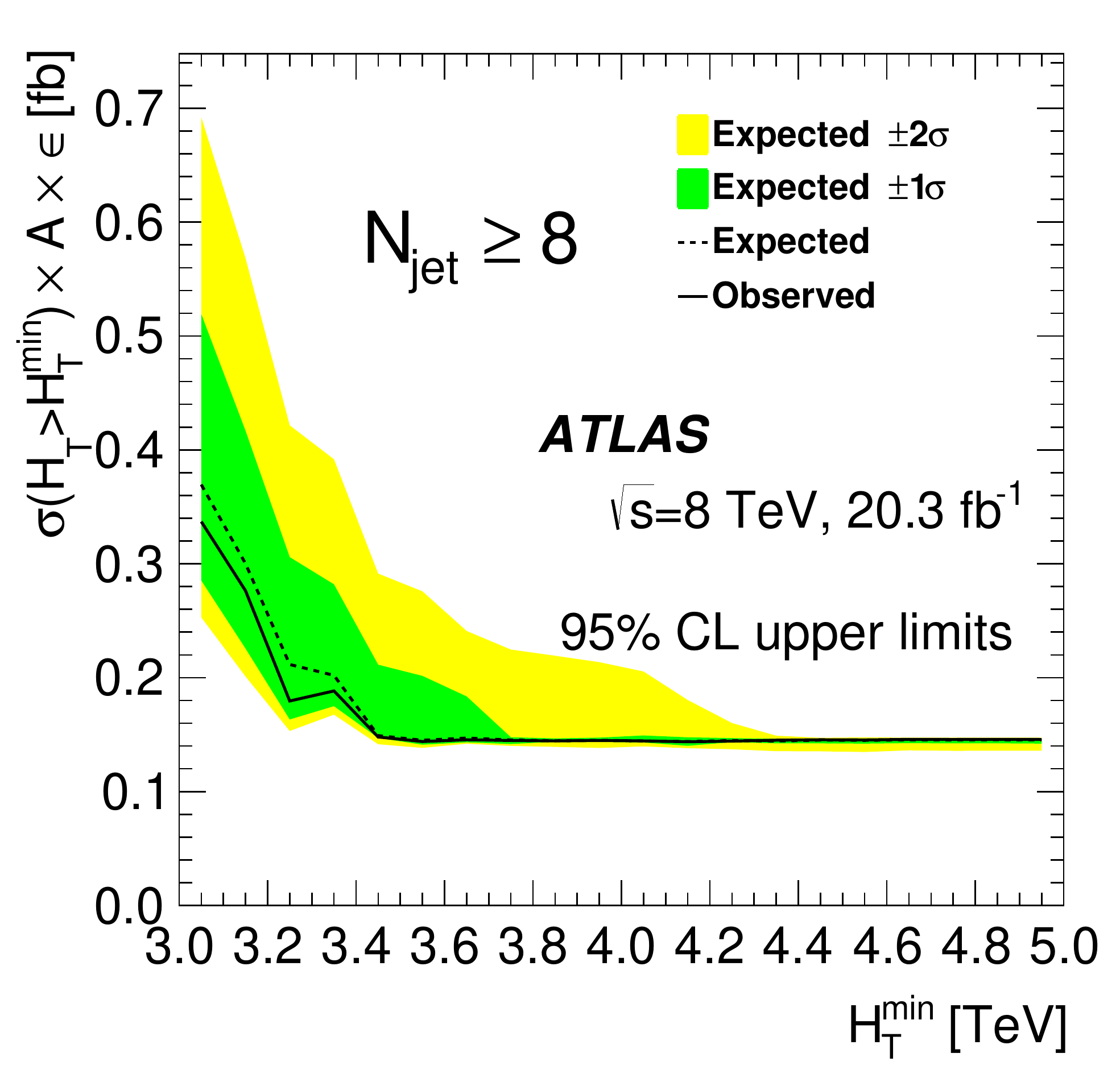}
\caption{Upper limits on the visible cross section (cross section
($\sigma$) times acceptance ($A$) times efficiency ($\epsilon$)) at the
95\% CL versus inclusive \HT{} for different inclusive jet multiplicities.
The solid (dashed) lines correspond to the observed (expected) upper
limits.
The green (dark) and yellow (light) bands represent one and two standard
deviations from the expected limits.}
\label{fig5}
\end{center}
\end{figure}

Using the observed number of events in each signal region compared to
the predicted number of SM background events,
model-independent limits on the observation of new phenomena at high
\HT{} and inclusive \NJ{} are set.
In addition, model-dependent limits on several black hole and string
ball signal models are set using the previous information, and estimates
of the acceptance and efficiency for each model.
The statistical procedure employed uses the one-sided profile-likelihood
test statistic~\cite{Cowan:2010js}.
The uncertainties are modelled with a convolution of Gaussian
probability density functions describing the uncertainties on the signal
or on the background.
The upper limits are derived from pseudo-experiments.

Counting experiments with $\HT > \HTinc$ as a function of \HTinc{} are
performed for each inclusive jet multiplicity $\NJ\ge 3$ to $\NJ\ge 8$.
Using the number of data events, estimated background, estimated
uncertainty on the background, luminosity, and uncertainty on the
luminosity, upper limits are calculated on the number of events divided
by the integrated luminosity.
Upper limits on the visible cross section (number of events divided by the
luminosity or cross section times acceptance times reconstruction
efficiency) at the 95\%~CL are shown in figure~\ref{fig5}.

The expected uncertainty bands narrow at high \HT{} as the test
statistic becomes discretely distributed due to the extremely small
background prediction.
These model-independent limits on the cross section times acceptance
times efficiency are as low as 0.14\,fb at the 95\% CL for minimum \HT{}
values above $4.3\TeV$ where no data events are observed.

To set limits on various models, the visible cross sections are divided
by the reconstruction efficiencies to obtain limits on fiducial cross
sections defined at the particle level.
The reconstruction efficiencies are calculated by taking the total
signal efficiency times acceptance and dividing by the fiducial
acceptance.
The fiducial acceptance at the generator particle level\footnote{This
includes parton showering and jet clustering, using the anti-$k_t$
algorithm with $R = 0.4$ on stable particles ($c\tau > 10$~mm),
including electrons and photons, but excluding muons and neutrinos.} is
defined from the simulated signal events with final states that pass the
jet $\pT > 50\GeV$ and $|\eta| < 2.8$ requirements.
The events are then counted in the different inclusive \NJ{} and \HT{}
regions.
Detector resolution can cause migration of signal events to different
regions of \NJ{} or \HT.
This can cause the reconstruction efficiency to be larger than unity in
some (\NJ, \HT) regions.

\begin{figure}[htb]
\begin{center}
\includegraphics[width=0.49\textwidth]{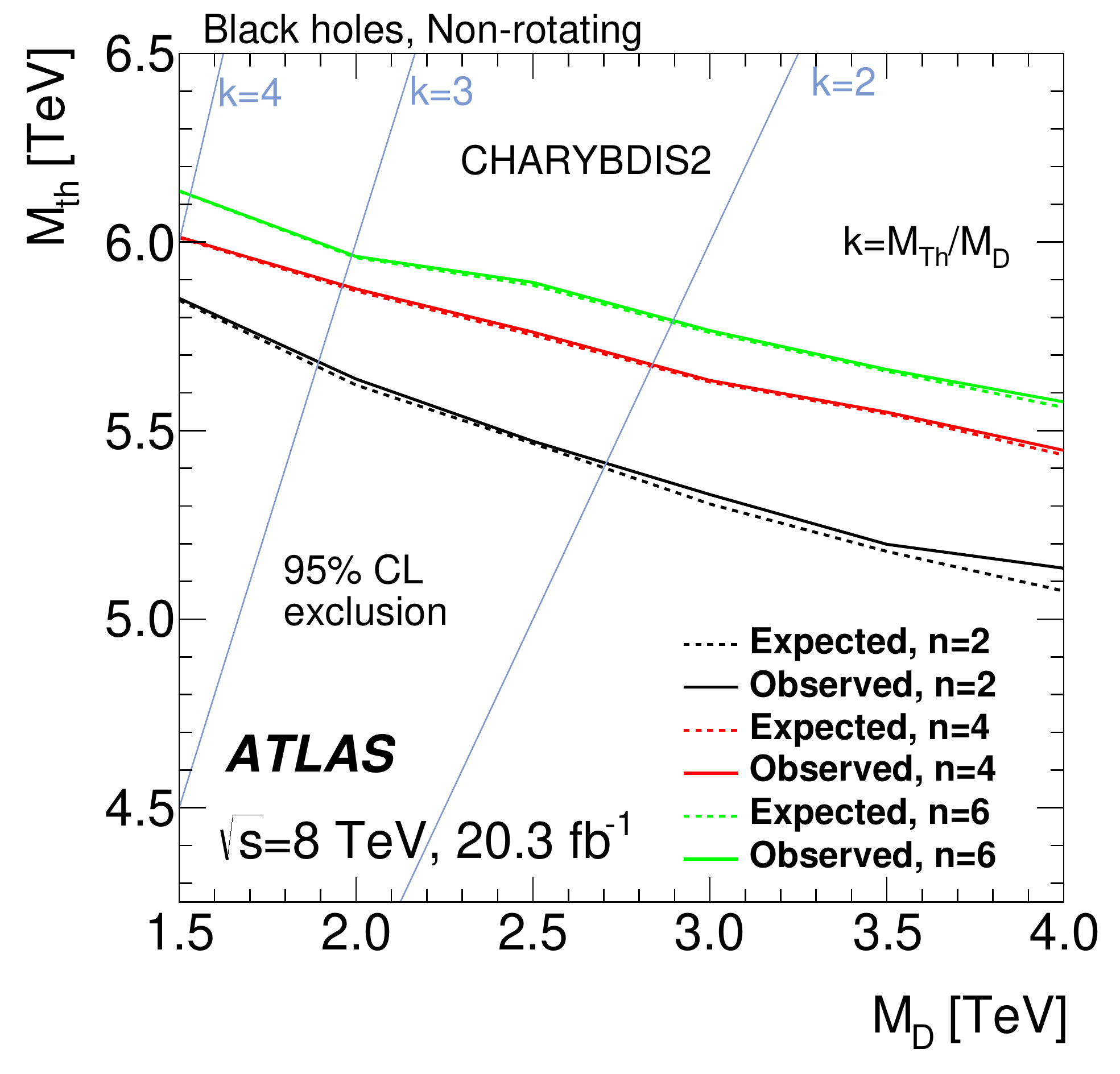}
\includegraphics[width=0.49\textwidth]{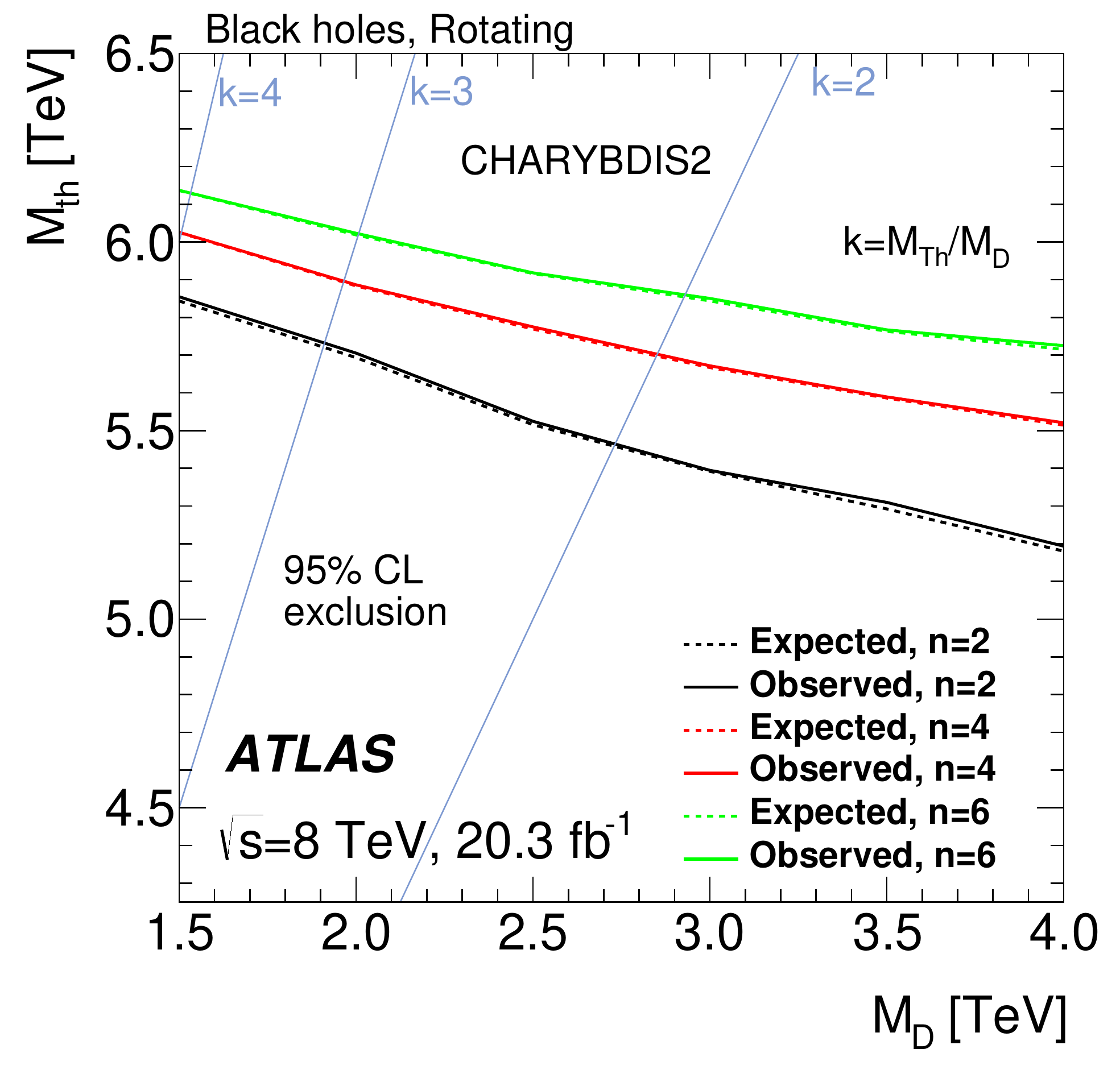}
\includegraphics[width=0.49\textwidth]{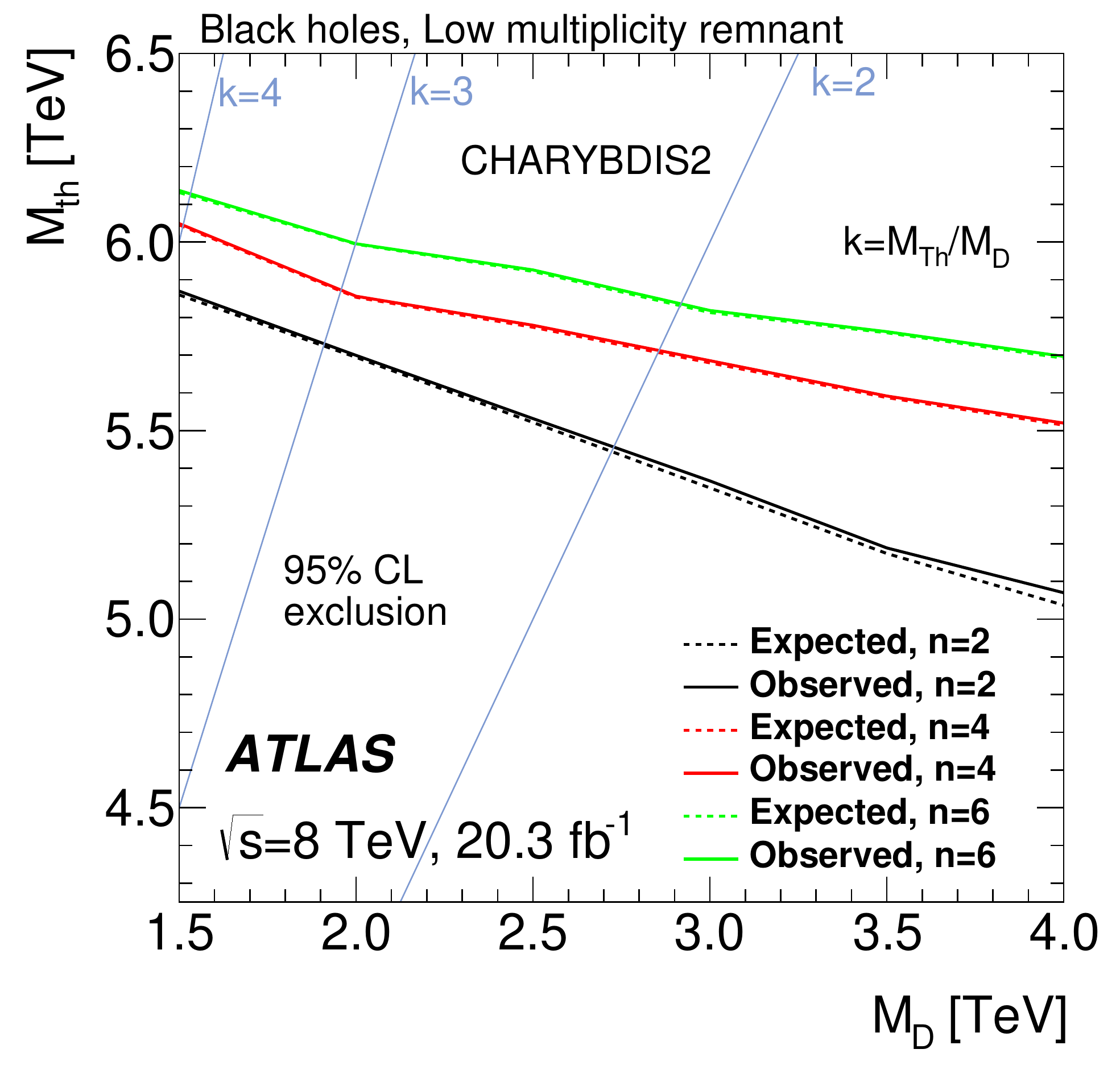}
\includegraphics[width=0.49\textwidth]{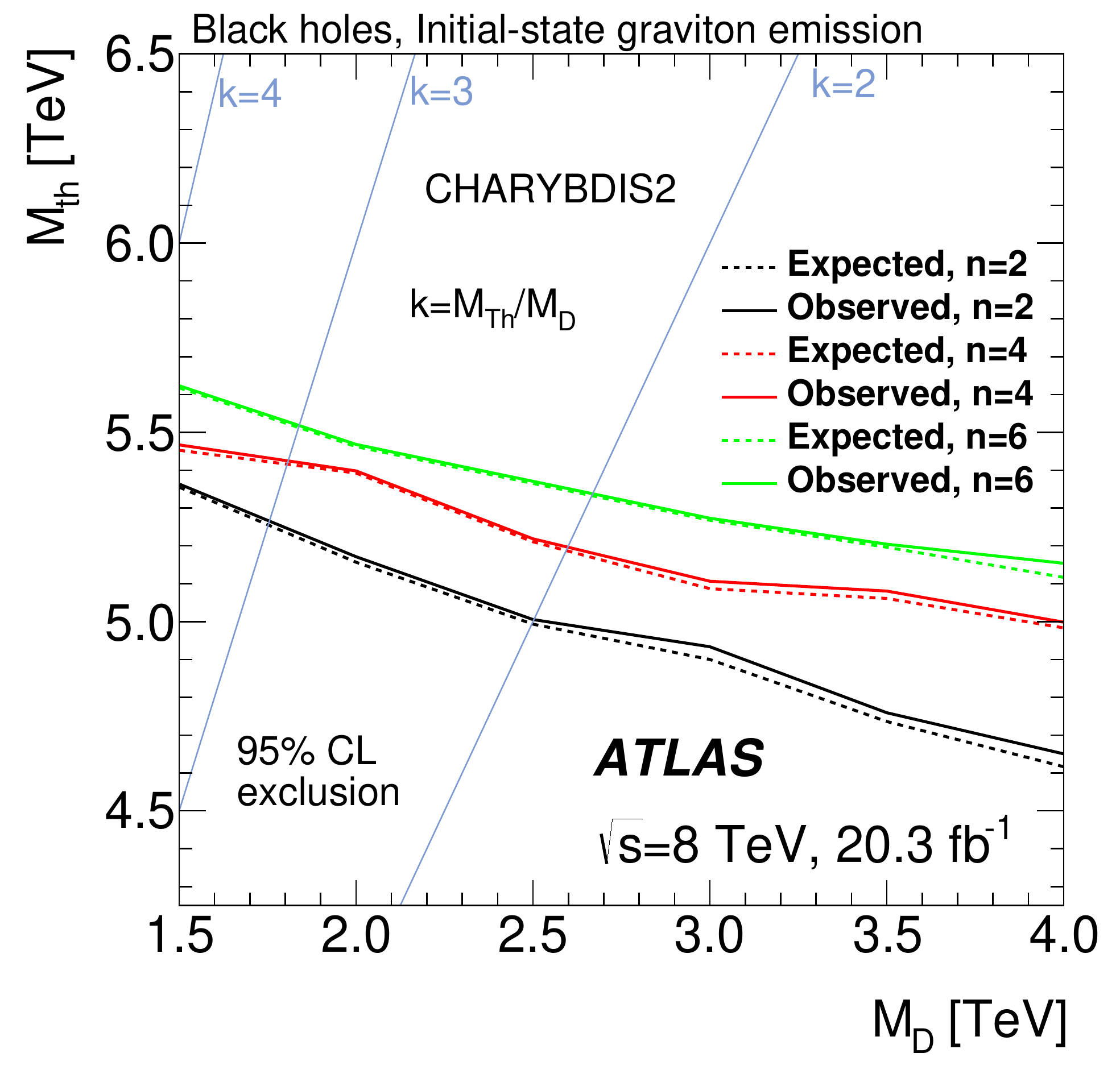}
\end{center}
\caption{Exclusion contours in the \Mth--\MD{} plane for different black
hole models in two, four, and six extra dimensions simulated with \charybdis.
The solid (dashed) lines show the observed (expected) 95\%~CL limits.
Masses below the corresponding lines are excluded.
Lines of fixed \Mth/\MD{} (defined as $k$) are shown.
The assumptions of the models are valid for $k \gg 1$.}
\label{fig8}
\end{figure}

\begin{figure}[htb]
\begin{center}
\includegraphics[width=0.49\textwidth]{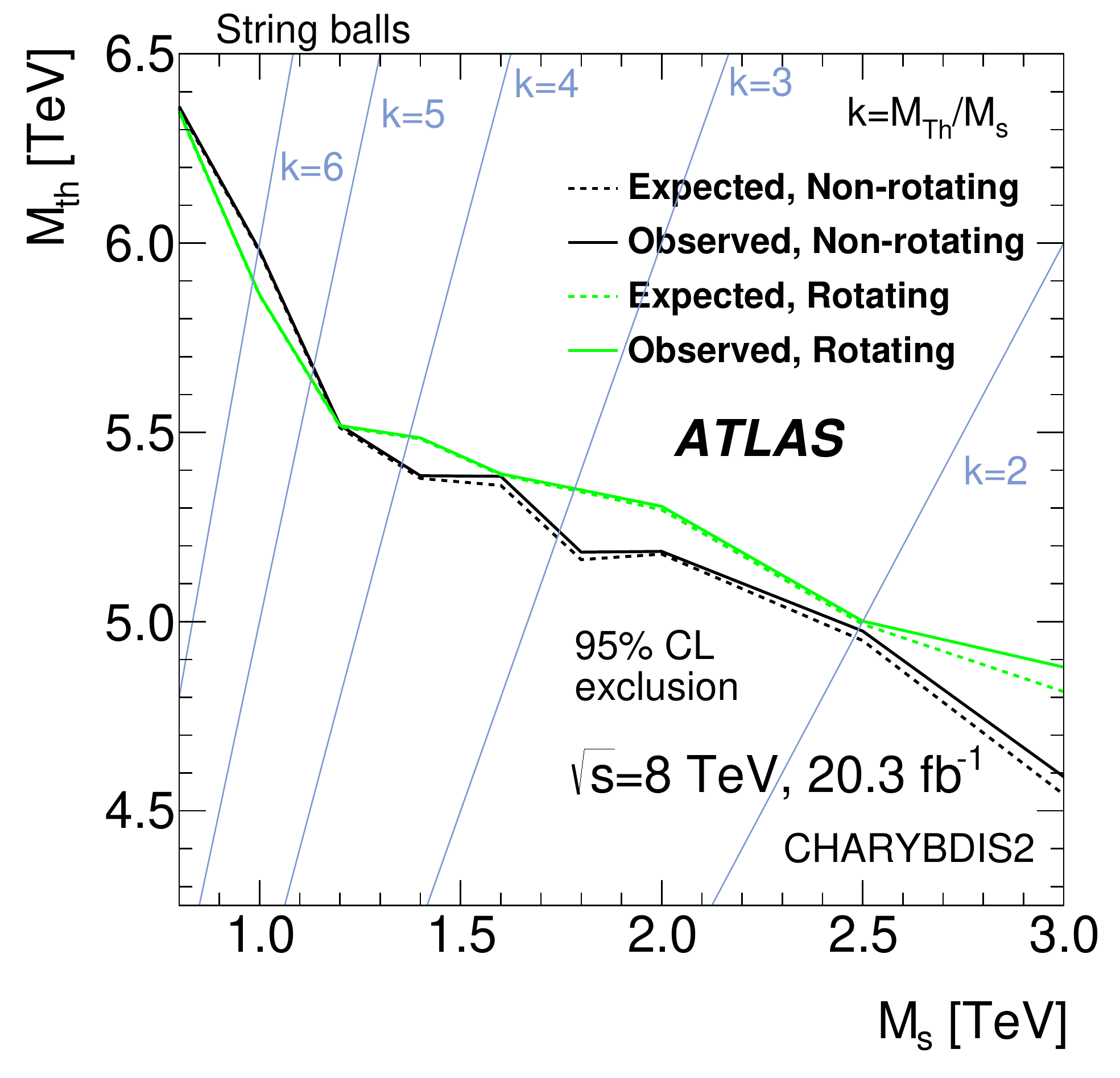}
\end{center}
\caption{Exclusion contours in the \Mth--$M_\mathrm{s}$ plane for
non-rotating and rotating string ball models simulated with \charybdis.
The solid (dashed) lines show the observed (expected) 95\%~CL limits.
Masses below the corresponding lines are excluded.
Lines of fixed \Mth/$M_\mathrm{s}$ (defined as $k$) are shown.
The assumptions of the models are valid for $k \gg 1$.}
\label{fig9}
\end{figure}

\begin{figure}[htb]
\begin{center}
\includegraphics[width=0.49\textwidth]{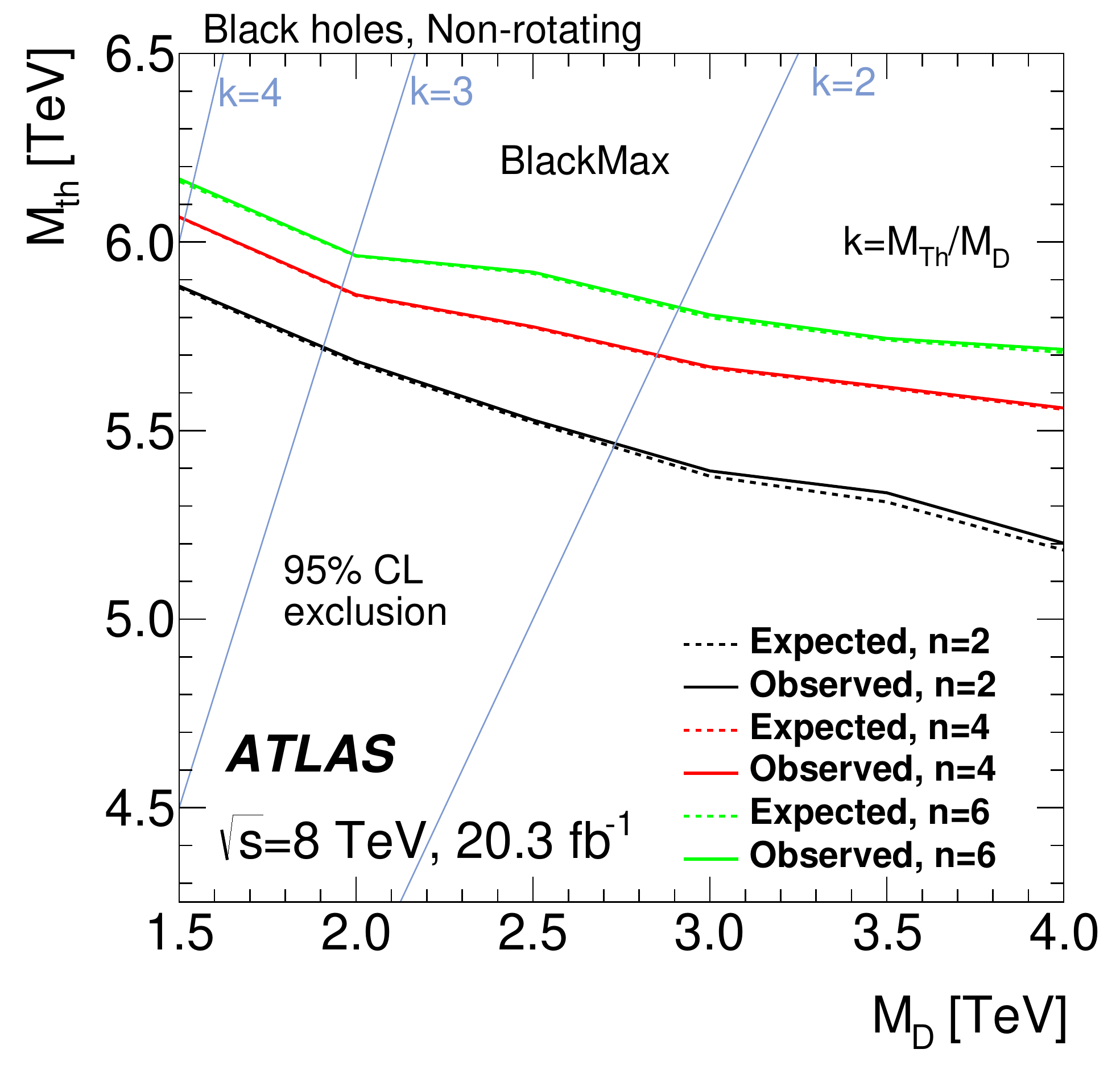}
\includegraphics[width=0.49\textwidth]{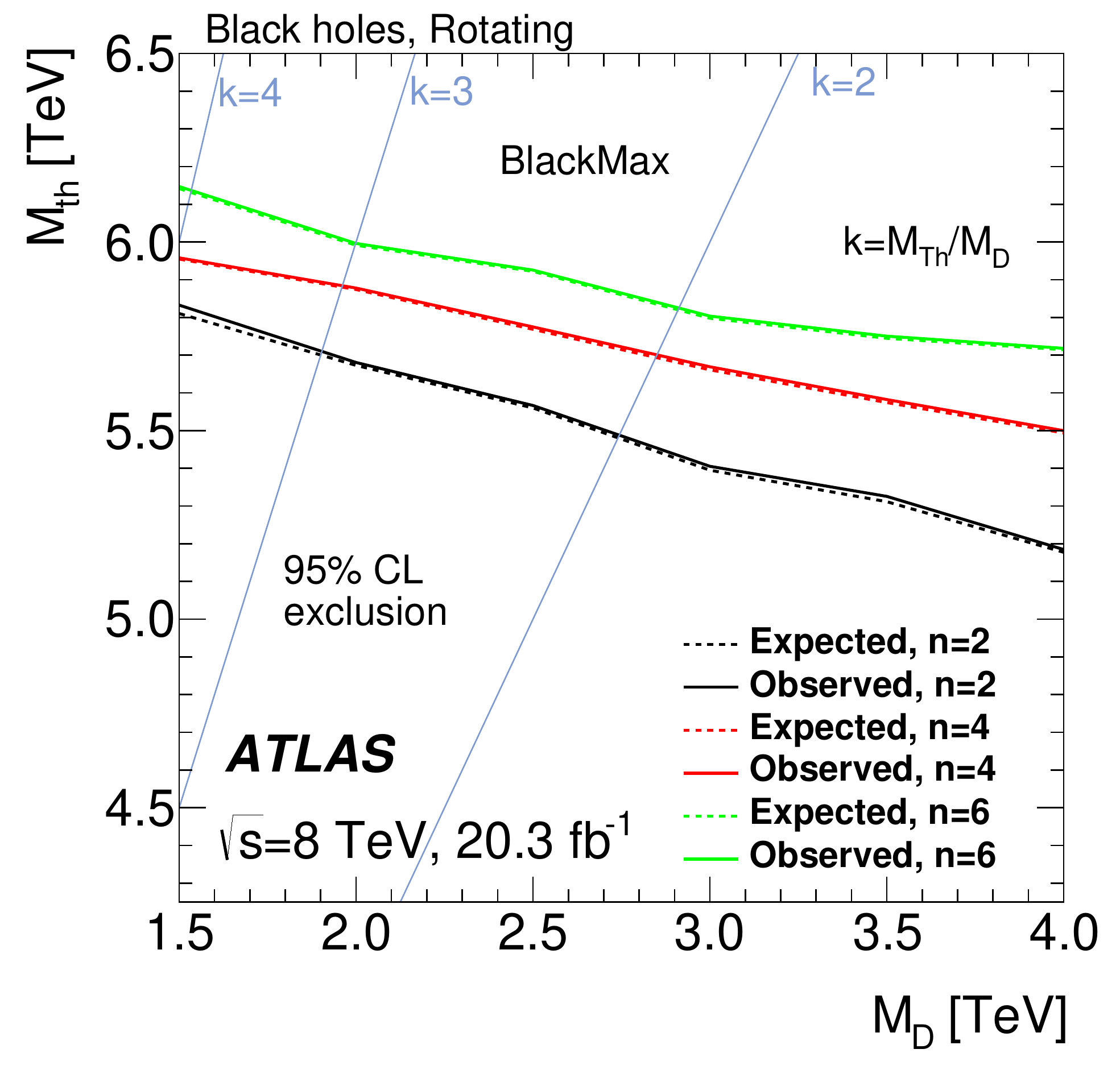}
\includegraphics[width=0.49\textwidth]{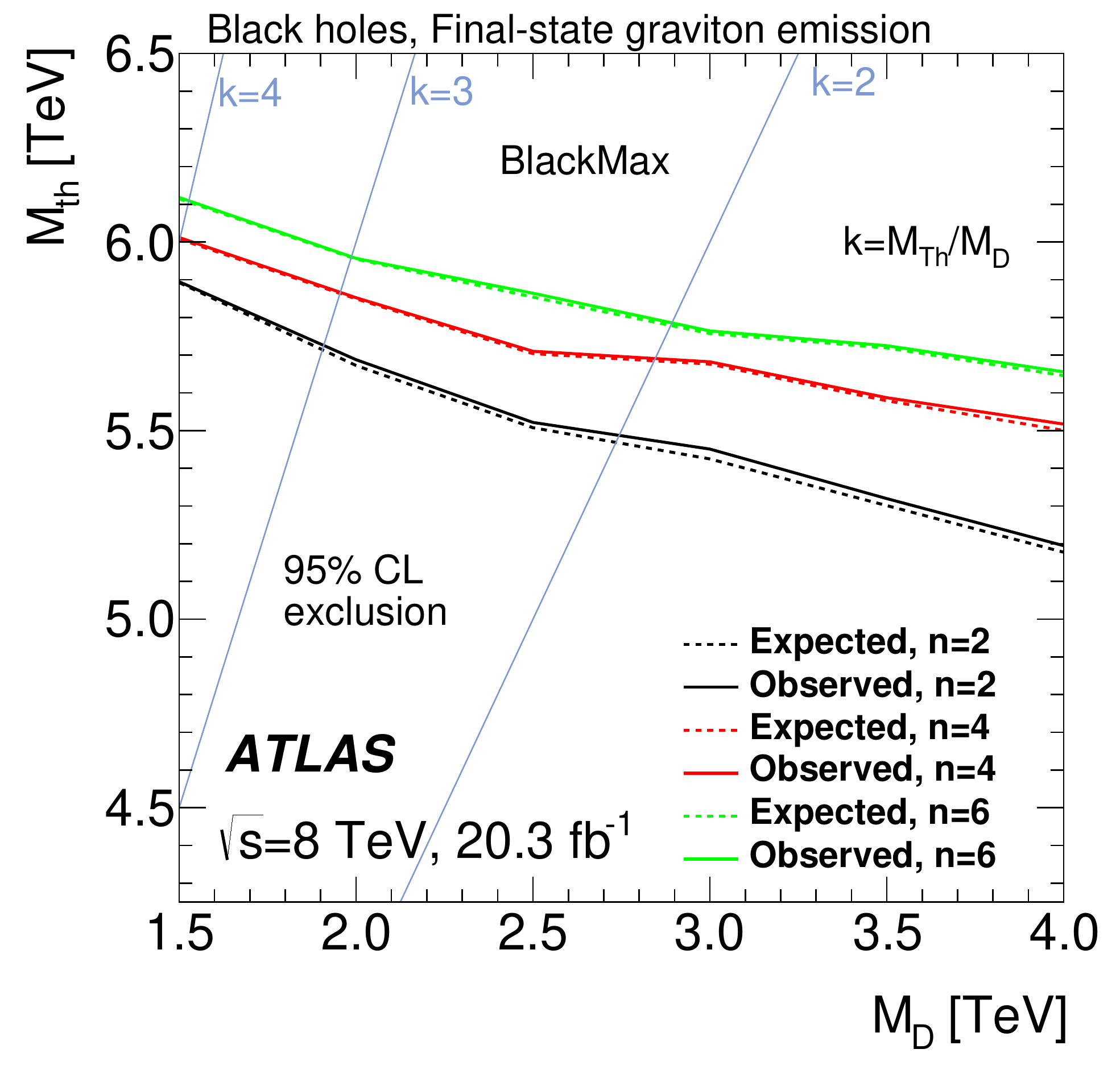}
\includegraphics[width=0.49\textwidth]{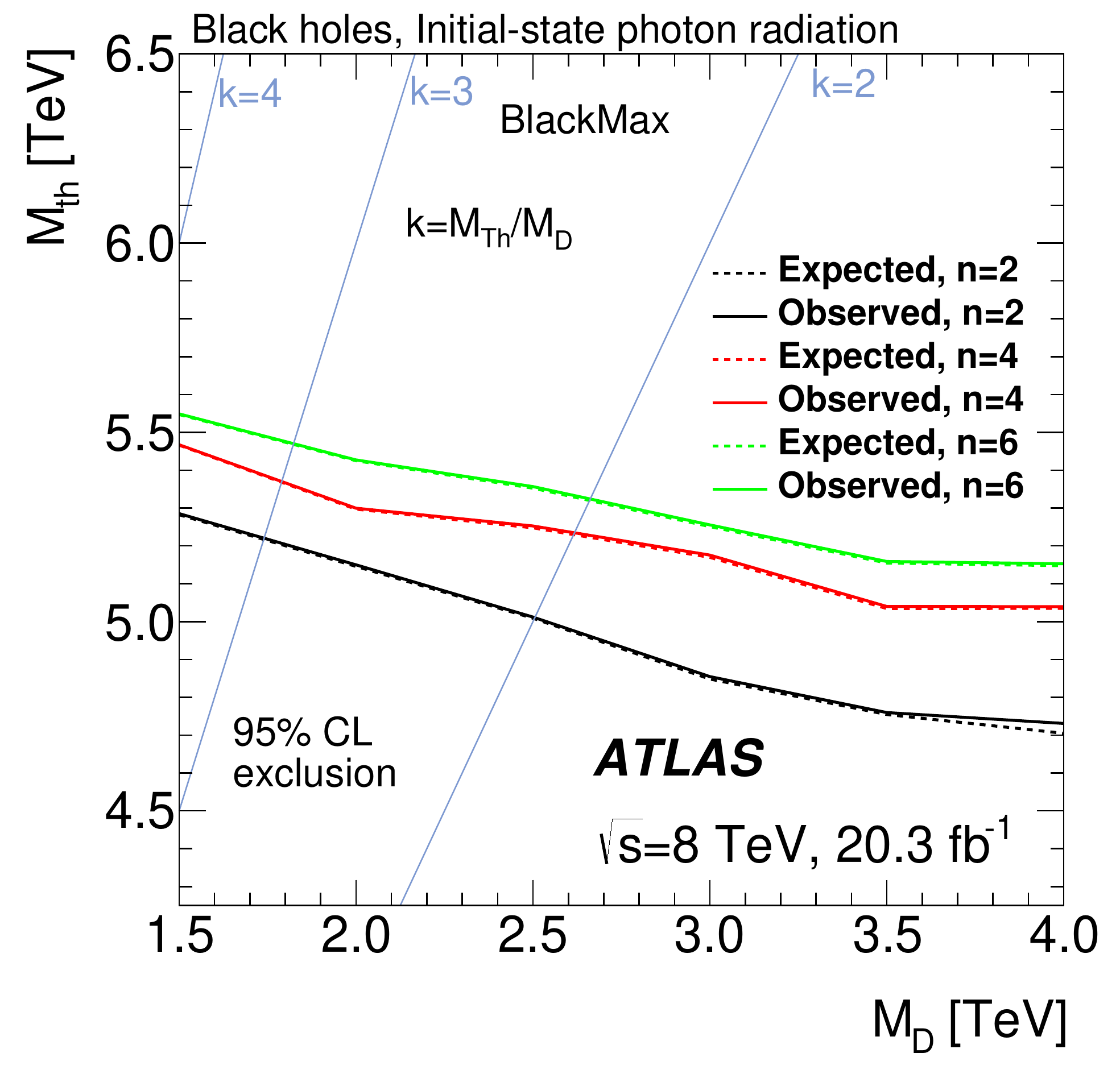}
\end{center}
\caption{Exclusion contours in the \Mth--\MD{} plane for different black
hole models in two, four, and six extra dimensions simulated with \blackmax.
The solid (dashed) lines show the observed (expected) 95\%~CL limits.
Masses below the corresponding lines are excluded.
Lines of fixed \Mth/\MD{} (defined as $k$) are shown.
The assumptions of the models are valid for $k \gg 1$.}
\label{fig10}
\end{figure}

The reconstruction efficiency depends on the particular black hole signal
production model and kinematic region $(\NJ, \HTinc)$ of interest.
The efficiencies are determined over the range of $n$, $\MD$, and $\Mth$
shown in figures~\ref{fig8}--\ref{fig10}.
The efficiencies are practically independent of the number of extra
dimensions in the model.
The mean reconstruction efficiencies are about 88\% with a variation
between models of about 1\%.
The RMS spread in efficiencies for a particular model is about 4\%
with a variation between models of about 1\%.

The limit on the cross section times acceptance is obtained from the
visible cross section and the average reconstruction efficiency from
many models, parameters, and kinematic properties.
The limit on the cross section times acceptance is as low as 0.16\,fb
at the 95\%~CL for minimum \HT{} values above about $4.3\TeV$.
This is comparable to the results in ref.~\cite{Chatrchyan:2013xva} in
which missing transverse momentum is included in the definition of \HT{}
if it is greater than $50\GeV$.
The results presented here are also compared to another ATLAS
analysis~\cite{Aad:2014gka}, in which a high-\pT{} lepton was required.
The intersection of these limits with theoretical predictions for the
cross section within the fiducial selections used in this analysis could
be used to constrain other models of new physics resulting in energetic
multi-jet final states.

Exclusion contours are obtained in the plane of \MD{} and \Mth{} for several
models.
For this purpose, counting experiments are performed to set a 95\%~CL
cross-section upper limit for each signal model and each $(\Mth, \MD)$
value of that signal model in this analysis.
In setting these limits, both the estimated background uncertainty and
signal uncertainties are taken into account.
The background and its uncertainty are the same as described previously
for the model-independent limits.
The uncertainties on the signal include the uncertainty due to the jet
energy scale and jet energy resolution, the statistical uncertainty on
the signal MC samples, and the uncertainty on the integrated
luminosity.
As the cross section for black hole production is known only
approximately and is highly model dependent, no theoretical uncertainty
on the signal cross section is applied.

For each grid point in the \Mth-\MD{} plane, the signal region which
gives the lowest expected $p$-value is used.
The most sensitive signal region for a particular signal model follows the
kinematics of the signal model.
For example, high-multiplicity signal regions are best for
high-multiplicity signal samples, and high-\HTinc{} regions are best for
high-\Mth{} signal samples.
Observed and expected exclusion contours for different \charybdis black
hole models are shown in figure~\ref{fig8}.
The observed and expected exclusions for $n=2$, $n=4$, and $n=6$ are shown.
In each exclusion figure, lines of fixed \Mth/\MD{} (defined as $k$) are
shown.
The assumptions of the models are not valid for $k=1$, but are valid for $k
\gg 1$.
These lines therefore form useful guidelines as to the validity of the models
across the plane.
The results for non-rotating and rotating string balls are shown in
figure~\ref{fig9}.

The exclusions tend to be stronger for higher $n$, due to the larger signal
cross sections.
For low values of \Mth/\MD, where there are the fewest
Hawking emissions (and where the semi-classical production assumptions are
least valid), the limits worsen.
The exclusions for non-rotating and rotating black hole models, with all
other parameters identical, appear similar.
Including initial-state radiation reduces the cross section and hence
the (\Mth, \MD) exclusion reach.
A low-multiplicity remnant state weakens the exclusion reach for $n=2$
at low values of \Mth/\MD, due to the reduced number of jets.
For string balls, the exclusion for the rotating case is similar to the
non-rotating case, in contrast to the result in ref.~\cite{Aad:2014gka}.

The exclusion contours for \blackmax models are shown in
figure~\ref{fig10}.
They show the same general features as the ones obtained with samples
generated by \charybdis.
\blackmax uses a final-burst remnant model, which gives
high-multiplicity remnant states~\cite{Dai:2007ki}.
Comparing non-rotating and rotating \charybdis and \blackmax results,
shows this analysis is insensitive to the different remnant models.
The results for the \blackmax{} model of production losses to photons is
comparable to the results for the \charybdis{} model of production
losses to gravitons.
Graviton emission in non-rotating black hole models weakens the exclusion
slightly, as a greater number of decay products carry missing energy and
do not contribute to the number of jets or \HT.

Contour limits of \Mth{} versus \MD{} are presented for a variety of
models.
These limits can be interpreted in terms of lower-mass limits on black
hole and string ball masses that range from 4.6 to $6.2\TeV$.
This is again comparable to the results in
ref.~\cite{Chatrchyan:2013xva} with the limits here being about $0.1\TeV$
higher in mass.
The results presented here are also compared with those of
ref.~\cite{Aad:2014gka}.
In the low-\MD{} region the results are comparable, while in the
high-\MD{} region the results presented here are a significant
improvement over those in ref.~\cite{Aad:2014gka}.
The latter analysis is affected by a significant loss in sensitivity for
the cases of rotating black holes and string balls, while the results
presented here, and those in ref.~\cite{Chatrchyan:2013xva}, are rather
independent of rotation.

\FloatBarrier

\section{Conclusion}
\label{sec:conclusion}

The production of events with multiple high-transverse-momentum jets is
measured using \lumi{} of proton--proton collision data recorded
at $\sqrt{s} = 8\TeV$ with the ATLAS detector at the LHC.
No significant excess beyond SM expectations is observed, and upper
limits on the visible cross sections for non-SM production of these final
states are set.
Using models for black hole and string ball production and decay,
exclusion contours are determined as a function of mass threshold and
the fundamental Planck scale.

The limit on the cross section times acceptance can be obtained from the
visible cross section and the average reconstruction efficiency taken
over many models, parameters, and kinematics.
The limit on the cross section times acceptance is as low as 0.16\,fb
at the 95\%~CL for minimum \HT{} values above about $4.3\TeV$.

Contour limits of \Mth{} versus \MD{} are presented for a variety of
models.
These limits can be interpreted in terms of lower-mass limits on black
hole and string ball masses that range from 4.6 to $6.2\TeV$.

\section*{Acknowledgements}

We thank CERN for the very successful operation of the LHC, as well as the
support staff from our institutions without whom ATLAS could not be
operated efficiently.

We acknowledge the support of ANPCyT, Argentina; YerPhI, Armenia; ARC,
Australia; BMWFW and FWF, Austria; ANAS, Azerbaijan; SSTC, Belarus; CNPq and FAPESP,
Brazil; NSERC, NRC and CFI, Canada; CERN; CONICYT, Chile; CAS, MOST and NSFC,
China; COLCIENCIAS, Colombia; MSMT CR, MPO CR and VSC CR, Czech Republic;
DNRF, DNSRC and Lundbeck Foundation, Denmark; EPLANET, ERC and NSRF, European Union;
IN2P3-CNRS, CEA-DSM/IRFU, France; GNSF, Georgia; BMBF, DFG, HGF, MPG and AvH
Foundation, Germany; GSRT and NSRF, Greece; RGC, Hong Kong SAR, China; ISF, MINERVA, GIF, I-CORE and Benoziyo Center, Israel; INFN, Italy; MEXT and JSPS, Japan; CNRST, Morocco; FOM and NWO, Netherlands; BRF and RCN, Norway; MNiSW and NCN, Poland; GRICES and FCT, Portugal; MNE/IFA, Romania; MES of Russia and NRC KI, Russian Federation; JINR; MSTD,
Serbia; MSSR, Slovakia; ARRS and MIZ\v{S}, Slovenia; DST/NRF, South Africa;
MINECO, Spain; SRC and Wallenberg Foundation, Sweden; SER, SNSF and Cantons of
Bern and Geneva, Switzerland; NSC, Taiwan; TAEK, Turkey; STFC, the Royal
Society and Leverhulme Trust, United Kingdom; DOE and NSF, United States of
America.

The crucial computing support from all WLCG partners is acknowledged
gratefully, in particular from CERN and the ATLAS Tier-1 facilities at
TRIUMF (Canada), NDGF (Denmark, Norway, Sweden), CC-IN2P3 (France),
KIT/GridKA (Germany), INFN-CNAF (Italy), NL-T1 (Netherlands), PIC (Spain),
ASGC (Taiwan), RAL (UK) and BNL (USA) and in the Tier-2 facilities
worldwide.

\bibliographystyle{elsarticle-num}
\bibliography{multijet}

\newpage
\begin{flushleft}
{\Large The ATLAS Collaboration}

\bigskip

G.~Aad$^{\rm 85}$,
B.~Abbott$^{\rm 113}$,
J.~Abdallah$^{\rm 151}$,
O.~Abdinov$^{\rm 11}$,
R.~Aben$^{\rm 107}$,
M.~Abolins$^{\rm 90}$,
O.S.~AbouZeid$^{\rm 158}$,
H.~Abramowicz$^{\rm 153}$,
H.~Abreu$^{\rm 152}$,
R.~Abreu$^{\rm 30}$
% ...
\bigskip
\\
% ...
$^{11}$ Institute of Physics, Azerbaijan Academy of Sciences, Baku, Azerbaijan\\
% ...
$^{30}$ CERN, Geneva, Switzerland\\
% ...
$^{85}$ CPPM, Aix-Marseille Universit{\'e} and CNRS/IN2P3, Marseille, France\\
% ...
$^{90}$ Department of Physics and Astronomy, Michigan State University, East Lansing MI, United States of America\\
% ...
$^{107}$ Nikhef National Institute for Subatomic Physics and University of Amsterdam, Amsterdam, Netherlands\\
% ...
$^{113}$ Homer L. Dodge Department of Physics and Astronomy, University of Oklahoma, Norman OK, United States of America\\
% ...
$^{151}$ Institute of Physics, Academia Sinica, Taipei, Taiwan\\
$^{152}$ Department of Physics, Technion: Israel Institute of Technology, Haifa, Israel\\
$^{153}$ Raymond and Beverly Sackler School of Physics and Astronomy, Tel Aviv University, Tel Aviv, Israel\\
% ...
$^{158}$ Department of Physics, University of Toronto, Toronto ON, Canada\\
% ...
\end{flushleft}

\end{document}